\documentclass[floatfix,twocolumn,showpacs,preprintnumbers,amsmath,amssymb,pra,superscriptaddress,longbibliography]{revtex4-2}
\usepackage{color}
\usepackage[usenames,dvipsnames,svgnames,table]{xcolor}
\usepackage[colorlinks=true,linkcolor=blue,urlcolor=blue,citecolor=blue]{hyperref}
\usepackage{mathtools}
\usepackage{graphicx}
\usepackage{dcolumn}
\usepackage{array}
\usepackage{lipsum}
\usepackage{bm}

\usepackage{amssymb}
\usepackage{multirow}
\usepackage{tabularx}
\usepackage{amsmath}
\usepackage{braket}
\usepackage{csquotes}
\graphicspath{{plots/}}
 \usepackage{lipsum}
\usepackage{mathrsfs}
\usepackage{MnSymbol}
\newcommand{\beq}{\begin{equation}}
\newcommand{\eeq}{\end{equation}}
\newcommand{\bea}{\begin{eqnarray}}
\newcommand{\eea}{\end{eqnarray}}

\begin{document}
\title{Second roton feature in the strongly coupled electron liquid}
\author{Thomas M.~Chuna}
\email{t.chuna@hzdr.de}
\affiliation{Center for Advanced Systems Understanding (CASUS), D-02826 G\"orlitz, Germany}
\affiliation{Helmholtz-Zentrum Dresden-Rossendorf (HZDR), D-01328 Dresden, Germany}

\author{Jan Vorberger}
\affiliation{Helmholtz-Zentrum Dresden-Rossendorf (HZDR), D-01328 Dresden, Germany}

\author{Panagiotis Tolias}
\affiliation{Space and Plasma Physics, Royal Institute of Technology (KTH), Stockholm, SE-100 44, Sweden}

\author{Alexander Benedix Robles}
\affiliation{Center for Advanced Systems Understanding (CASUS), D-02826 G\"orlitz, Germany}
\affiliation{Helmholtz-Zentrum Dresden-Rossendorf (HZDR), D-01328 Dresden, Germany}
\affiliation{Technische  Universit\"at  Dresden,  D-01062  Dresden,  Germany}

\author{Michael Hecht}
\affiliation{Center for Advanced Systems Understanding (CASUS), D-02826 G\"orlitz, Germany}
\affiliation{Helmholtz-Zentrum Dresden-Rossendorf (HZDR), D-01328 Dresden, Germany}

\author{Phil-Alexander Hofmann}
\affiliation{Center for Advanced Systems Understanding (CASUS), D-02826 G\"orlitz, Germany}
\affiliation{Helmholtz-Zentrum Dresden-Rossendorf (HZDR), D-01328 Dresden, Germany}
\affiliation{Technische  Universit\"at  Dresden,  D-01062  Dresden,  Germany}

\author{Zhandos A.~Moldabekov}
\affiliation{Center for Advanced Systems Understanding (CASUS), D-02826 G\"orlitz, Germany}
\affiliation{Helmholtz-Zentrum Dresden-Rossendorf (HZDR), D-01328 Dresden, Germany}

\author{Tobias Dornheim}
\email{t.dornheim@hzdr.de}
\affiliation{Center for Advanced Systems Understanding (CASUS), D-02826 G\"orlitz, Germany}
\affiliation{Helmholtz-Zentrum Dresden-Rossendorf (HZDR), D-01328 Dresden, Germany}

\begin{abstract}
We present extensive \emph{ab initio} path integral Monte Carlo (PIMC) results for the dynamic properties of the finite temperature uniform electron gas (UEG) over a broad range of densities, $2\leq r_s\leq300$. We demonstrate that the direct analysis of the imaginary-time density--density correlation function (ITCF) allows for a rigorous assessment of the density and temperature dependence of the previously reported roton-type feature [T.~Dornheim, \emph{Phys.~Rev.~Lett.}~\textbf{121}, 255001 (2018)] at intermediate wavenumbers. We clearly resolve the emergence of a second roton at the second harmonic of the original feature for $r_s\gtrsim100$, which we identify as an incipient phonon dispersion. Finally, we use our highly accurate PIMC results for the ITCF as the basis for an analytic continuation to compute the dynamic structure factor, which additionally substantiates the existence of the second roton in the strongly coupled electron liquid. Our investigation further elucidates the complex interplay between quantum delocalization and Coulomb coupling in the UEG. All PIMC results are freely available online and provide valuable benchmarks for other theoretical methodologies and approximations.
\end{abstract}
\maketitle

\section{Introduction}

The uniform electron gas (UEG)~\cite{quantum_theory,loos,review}---also known as jellium and quantum one-component plasma (OCP) in the literature---constitutes one of the most important and most widely studied model systems in physics, quantum chemistry and related disciplines. By treating the positively charged nuclei as a rigid homogeneous neutralizing background, the UEG model allows one to focus on electronic exchange--correlation (XC) effects, making it the archetypal model for interacting electrons with a broad range of applications. Despite its seeming simplicity, the accurate description of its properties is highly non-trivial and requires sophisticated numerical methods. At ambient conditions, the electrons are assumed to be in their respective ground state (i.e., in the limit of $T=0$), and accurate results for a broad range of observables, including the energy~\cite{Ceperley_Alder_PRL_1980,Ortiz_PRB_1994,Ortiz_PRL_1999,Spink_PRB_2013,Shepherd_JCP_2012} and the linear density response~\cite{moroni,moroni2,bowen2,Chen2019}, are available from zero-temperature quantum Monte Carlo (QMC) simulations~\cite{Foulkes_RMP_2001,anderson2007quantum}.

Over the last decade, there has emerged a surge of interest into the properties of the UEG at finite temperatures, with a particular focus on the warm dense matter (WDM) regime~\cite{review}. Specifically, WDM is usually defined by $r_s\sim\Theta\sim1$ (with $r_s$ the Wigner-Seitz radius and $\Theta=k_\textnormal{B}T/E_\textnormal{F}$ the reduced temperature with the electronic Fermi energy $E_\textnormal{F}$)~\cite{Ott2018}, which gives rise to the complex interplay of quantum delocalization and degeneracy, strong thermal excitations, and Coulomb interaction~\cite{new_POP,Dornheim_review,wdm_book}, see also a recent roadmap on WDM physics~\cite{vorberger2025roadmapwarmdensematter}. In nature, WDM occurs in a variety of astrophysical objects such as giant planet interiors~\cite{wdm_book,Benuzzi_Mounaix_2014,Guillot2018}, brown dwarfs~\cite{becker} and white dwarf atmospheres~\cite{Kritcher2020,SAUMON20221}. Furthermore, the fusion fuel and ablator material in inertial confinement fusion experiments~\cite{Zylstra2022,AbuShawareb_PRL_2024,drake2018high} have to traverse the WDM regime in a controlled way to reach ignition~\cite{hu_ICF}, making the accurate understanding of this regime paramount for the further development of inertial fusion energy into a clean and abundant source of energy~\cite{Batani_roadmap}. Naturally, the burgeoning interest in these applications has sparked a high demand for the accurate parametrization of the UEG at finite temperatures~\cite{review,dornheim_ML,ksdt,status,groth_prl,Dornheim_PRL_2020_ESA,Dornheim_PRB_ESA_2021} based on \emph{ab initio} path integral Monte Carlo (PIMC) simulations~\cite{Dornheim_JCP_2015,Brown_PRL_2013,Schoof_PRL_2015,dornheim_prl,Dornheim_JCP_xi_2023,dornheim_sign_problem,Dornheim_PRL_2020}.

A particularly intriguing regime is given by the strongly coupled electron liquid (loosely defined by $r_s\gtrsim10$ at $\Theta=1$)~\cite{dornheim_electron_liquid}, which allows one to isolate the interplay of strong Coulomb repulsion with both quantum delocalization and thermal excitation effects~\cite{Dornheim_EPL_2024,Dornheim_PRB_2024}. These conditions give rise to a non-monotonic dispersion of the dynamic structure factor (DSF) $S(\mathbf{q},\omega)$~\cite{dornheim_dynamic,Dornheim_Nature_2022}, which phenomenologically resembles the roton feature known from quantum liquids such as ultracold helium~\cite{Trigger,Ferre_PRB_2016,Godfrin2012,Dornheim_SciRep_2022} and also from the classical OCP~\cite{Mithen_AIP_2012,Kalman_2010}. This nontrivial effect has been explained from different but consistent points of view in terms of the alignment of pairs of electrons~\cite{Dornheim_Nature_2022,Dornheim_Force_2022} and as an excitonic mode~\cite{koskelo2023shortrange,Takada_PRB_2016}.

At finite temperatures, state-of-the-art PIMC simulations give one access to the imaginary-time density--density correlation function (ITCF)~\cite{Dornheim_JCP_ITCF_2021,Motta_JCP_2015,Boninsegni1996}
\begin{eqnarray}\label{eq:define_ITCF}
    F(\mathbf{q},\tau) = \braket{\hat{n}(\mathbf{q},0)\hat{n}(-\mathbf{q},\tau)}\ ,
\end{eqnarray}
which is connected to the DSF via a two-sided Laplace transform~\cite{JARRELL1996133}:
\begin{eqnarray}\label{eq:Laplace}
    F(\mathbf{q},\tau) = \mathcal{L}\left[S(\mathbf{q},\omega)\right] = \int_{-\infty}^\infty \textnormal{d}\omega\ S(\mathbf{q},\omega)\ e^{-\tau\omega}\ ;
\end{eqnarray}
note that Eq.~(\ref{eq:define_ITCF}) corresponds to the usual intermediate scattering function $F(\mathbf{q},t)$ but evaluated for an imaginary time $t=-i\tau$ with $\tau\in[0,\beta]$ where $\beta=1/k_\textnormal{B}T$ is the inverse temperature. The task at hand is then the numerical inversion of Eq.~(\ref{eq:Laplace}), which is a notoriously difficult and, in fact, exponentially ill-posed problem. Nevertheless, a gamut of methods have been introduced over the years, including maximum entropy solvers~\cite{JARRELL1996133, Fuchs_PRE_2010,Boninsegni_maximum_entropy,chuna2025dualformulationmaximumentropy}, genetic algorithms~\cite{Vitali_PRB_2010,Bertaina_AdvPhysX_2017}, as well as neural networks~\cite{Yao_2022,Yoon_PRB_2018,Fournier_PRL_2020}. For the specific case of the UEG, Dornheim and co-workers~\cite{dornheim_dynamic,dynamic_folgepaper,Dornheim_PRE_2020} have proposed to stochastically sample the dynamic local field correction [cf.~Eq.~(\ref{eq:define_G}) below], which allows one to incorporate a number of additional exact constraints and, in this way, to make the reconstruction of $S(\mathbf{q},\omega)$ tractable.
Available results from this procedure are currently limited to $r_s\leq20$. Very recently, Chuna \emph{et al.}~\cite{chuna2025estimatesdynamicstructurefactor} have combined the maximum entropy method with a data driven default model that is directly constructed from the PIMC results for $F(\mathbf{q},\tau)$ and applied it to the UEG at $\Theta=1$ over a broad range of densities, $2\leq r_s \leq 200$.
As it is expected, they have reported a pronounced roton minimum in the DSF around intermediate wavenumbers, which becomes substantially deeper for large values of $r_s$.

As an alternative to the difficult problem of analytic continuation, one might also consider directly analyzing the ITCF in the imaginary--time domain~\cite{Dornheim_MRE_2023,Dornheim_PTR_2023}. Indeed, owing to the uniqueness of the two-sided Laplace transform, $F(\mathbf{q},\tau)$ contains, by definition, exactly the same information as $S(\mathbf{q},\omega)$, only in an alternative representation. In fact, switching from $\omega$- to $\tau$-space can even be advantageous for the interpretation of X-ray Thomson scattering (XRTS) experiments~\cite{Dornheim_review,Dornheim_T_2022,Dornheim_T_follow_up,Dornheim_SciRep_2024,Dornheim_Science_2024,dornheim2024modelfreerayleighweightxray,shi2025firstprinciplesanalysiswarmdense,Schoerner_PRE_2023,Vorberger_PhysLettA_2024,Bellenbaum_APL_2025}, as it allows for a straightforward deconvolution of the physical information from the x-ray source function and from detector effects~\cite{Gawne_JAP_2024}. Furthermore, Dornheim \emph{et al.}~\cite{Dornheim_MRE_2023,Dornheim_moments_2023} have demonstrated that one can extract a wealth of information from the ITCF, including the roton feature at low densities.

In the present work, we further explore these ideas by presenting a detailed analysis of the ITCF of the finite-temperature UEG over a broad range of densities and temperatures. First, we show that the ITCF is indeed capable of revealing the roton feature at around twice the Fermi wavenumber, following the same trends with respect to $r_s$ as those observed for $S(\mathbf{q},\omega)$ via analytic continuation~\cite{chuna2025estimatesdynamicstructurefactor}. Second, we find an additional feature at the second harmonic of the original roton, which can be traced back to the emerging spatial order in the system at low densities. In contrast to the classical OCP, however, these high harmonics are strongly damped by quantum delocalization effects, which become increasingly important for large wavenumbers. Finally, we use the recent \texttt{PyLIT} code~\cite{pylit} to carry out an analytic continuation to $S(\mathbf{q},\omega)$, which further substantiates the existence of the second roton in the strongly coupled electron liquid.

The paper is organized as follows: In Sec.~\ref{sec:theory}, we provide the necessary theoretical background, including some details on the utilized PIMC implementation~(\ref{sec:PIMC}), linear response theory (\ref{sec:LRT}), the interpretation of the physics encoded into imaginary-time correlation functions (\ref{sec:ITCF}), and the analytic continuation to obtain the dynamic structure factor (\ref{sec:anal_cont_theory}). Sec.~\ref{sec:results} is devoted to the analysis and discussion of our PIMC simulation results in terms of the ITCF (\ref{sec:imaginary}) and in terms of the DSF (\ref{sec:anal_cont}). The paper is concluded by a summary and outlook in Sec.~\ref{sec:outlook}. The Appendix~\ref{sec:long_wavelength_derivation} features a detailed derivation of the exact long-wavelength limit of the ITCF.

\section{Theory\label{sec:theory}}

We assume Hartree atomic units throughout this work with the exception of the Appendix~\ref{sec:long_wavelength_derivation}.

\subsection{Path integral Monte Carlo\label{sec:PIMC}}

We use the direct PIMC method as it has been introduced in detail, e.g., in Refs.~\cite{cep,boninsegni1}. Since we do not impose any nodal restrictions~\cite{Ceperley1991}, our simulations are exact within the statistical Monte Carlo error bars for a given system size. Specifically, we explicitly sample all the positive and the negative contributions to the partition function~\cite{dornheim_sign_problem,Dornheim_permutation_cycles}, leading to average signs of $S\approx0.013$ ($r_s=2$) to $S\approx0.988$ ($r_s=100$) at $\Theta=1$. This makes our simulations computationally expensive (the required compute time scales as $\sim1/S^2$, leading to a ten thousand fold increase at $r_s=2$), but still feasible on modern high-performance computing systems. For $r_s\geq200$, we do not attempt to sample any permutations, which are suppressed by the strong Coulomb repulsion in this regime. An additional advantage of not imposing any nodal restrictions is that we retain full access to the imaginary-time structure of the simulated system, giving us straightforward access to the ITCF, Eq.~(\ref{eq:define_ITCF}) [as well as a host of other imaginary-time properties, including the Matsubara Green function~\cite{boninsegni1,boninsegni2,Filinov_PRA_2012} and higher-order ITCFs that are linked to nonlinear density response properties~\cite{Dornheim_JCP_ITCF_2021,Dornheim_review,Dornheim_CPP_2022}].

We use the \emph{primitive factorization} $e^{-\epsilon\hat{H}}\approx e^{-\epsilon\hat{K}}e^{-\epsilon\hat{V}}$, where $\epsilon=\beta/P$ is a discrete imaginary time step and $\hat{H}=\hat{K}+\hat{V}$ denotes the full Hamiltonian consisting of both kinetic and interaction contributions, which becomes exact in the limit of $P\to\infty$. In practice, we use $P=200$, which is sufficient to reduce factorization errors below the level of statistical noise (see also the recent Ref.~\cite{dornheim2025applicationsphericallyaveragedpair}), and also sufficient to resolve $F(\mathbf{q},\tau)$ on an appropriate $\tau$-grid.
For completeness, we note that higher order factorization schemes have been explored in the literature~\cite{cep,Takahashi_Imada_higher_order,sakkos_JCP_2009,Chin_PRE_2015,Dornheim_NJP_2015,Dornheim_CPP_2019,Zillich_JCP_2010}, but they are not required for the present conditions.

\subsection{Linear response theory\label{sec:LRT}}

Linear response theory is ubiquitous throughout warm dense matter theory~\cite{Dornheim_review}, and quantum many-body theory in general. The key property is given by the dynamic density response function~\cite{quantum_theory,kugler1}
\begin{eqnarray}\label{eq:define_G}
    \chi(\mathbf{q},\omega) = \frac{\chi_0(\mathbf{q},\omega)}{1 - \frac{4\pi}{q^2}\left[ 1 - G(\mathbf{q},\omega)\right]\chi_0(\mathbf{q},\omega)}\ ,
\end{eqnarray}
which describes the system's response to an external harmonic perturbation of wavevector $\mathbf{q}$ and frequency $\omega$; it is often conveniently expressed in terms of the finite-temperature Lindhard function $\chi_0(\mathbf{q},\omega)$, which is known semi-analytically~\cite{quantum_theory, tolias2024density} and describes the density response of the ideal Fermi gas at the same conditions, and the dynamic local field correction $G(\mathbf{q},\omega)$, which is generally unknown and contains the complete $\mathbf{q}$- and $\omega$-resolved information about electronic XC effects. As a consequence, $G(\mathbf{q},\omega)$ constitutes important input for a plethora of practical applications such as the interpretation of XRTS experiments~\cite{Fortmann_PRE_2010,Poole_PRR_2024,Dornheim_PRL_2020_ESA}, the exchange--correlation kernel for linear-response time-dependent density functional theory calculations~\cite{Ramakrishna_PRB_2021,Moldabekov_PRR_2023,ullrich2011time}, and the construction of advanced, non-local and explicitly thermal exchange--correlation functionals for thermal density functional theory~\cite{pribram}.

Setting $G(\mathbf{q},\omega)\equiv0$ in Eq.~(\ref{eq:define_G}) corresponds to the well-known \emph{random phase approximation} (RPA), which describes the density response on the mean-field level. A more sophisticated ansatz is given by the \emph{static approximation} $G(\mathbf{q},\omega)\equiv G(\mathbf{q},0)$, which has been shown to give highly accurate results for weak to moderate coupling, $r_s\lesssim5$~\cite{dornheim_dynamic,dynamic_folgepaper,Hamann_PRB_2020}. 
Further, the static local field correction is available as a highly accurate neural network representation~\cite{dornheim_ML} that combines finite-$T$ PIMC with ground state QMC results~\cite{moroni2,cdop} and covers the entire relevant parameter space. A convenient alternative is given by the \emph{effective static approximation} (ESA)~\cite{Dornheim_PRL_2020_ESA}, which combines PIMC results for $q\lesssim3q_\textnormal{F}$ with an appropriate short wavelength limit that is related to the pair correlation function~\cite{holas_limit,stls}, and which is also available as an analytical parametrization~\cite{Dornheim_PRB_ESA_2021}.

Finally, we mention that $G(\mathbf{q},\omega)$ [and its dynamic Matsubara equivalent $\widetilde{G}(\mathbf{q},z_l)$, where $z_l=i2\pi l/\beta$ are the discrete imaginary Matsubara frequencies~\cite{Tolias_JCP_2024,Dornheim_PRB_2024,Li_PRB_2025,IIT,stls}] is of central importance for the active field of dielectric theories, which deal with approximate closure relations for the static or dynamic local field correction on various levels of sophistication~\cite{stls,stls2,stls_original,vs_original,schweng,arora,castello2021classical,Tolias_JCP_2021,Tolias_JCP_2023,Tolias_PRB_2024,tanaka_hnc,Tanaka_CPP_2017,Tolias_CPP_2025}.

In the context of the present work, the most important application of linear response theory is given by the fluctuation--dissipation theorem~\cite{quantum_theory}
\begin{eqnarray}\label{eq:FDT}
S(\mathbf{q},\omega) = - \frac{\textnormal{Im}\chi(\mathbf{q},\omega)}{\pi n (1-e^{-\beta\omega})}\ ,
\end{eqnarray}
which relates $\chi(\mathbf{q},\omega)$ to the DSF and, via Eq.~(\ref{eq:Laplace}), also to the ITCF. In addition, we mention the imaginary-time version of this theorem~\cite{Dornheim_MRE_2023},
\begin{eqnarray}\label{eq:static_chi}
    \chi(\mathbf{q},0) = - n \int_0^\beta \textnormal{d}\tau\ F(\mathbf{q},\tau)\ ,
\end{eqnarray}
where $n=N/V$ is the number density. Eq.~(\ref{eq:static_chi}) directly implies that one can obtain the full wavenumber dependence of the static linear density response $\chi(\mathbf{q},0)$ from a single simulation of the unperturbed UEG by integrating over the ITCF. Inverting Eq.~(\ref{eq:define_G}) for $G(\mathbf{q},0)$,
\begin{eqnarray}\label{eq:G_inverted}
    G(\mathbf{q},0) = 1 + \frac{q^2}{4\pi}\left[
\frac{1}{\chi(\mathbf{q},0)} - \frac{1}{\chi_0(\mathbf{q},0)}
    \right]\, ,
\end{eqnarray}
then gives one access to $G(\mathbf{q},\omega)$ on the level of the static approximation~\cite{Dornheim_PRB_ESA_2021,chuna2025estimatesdynamicstructurefactor}.

\subsection{Imaginary-time correlation function\label{sec:ITCF}}

Extensive discussions of the physical content encoded into $F(\mathbf{q},\tau)$ have been presented in the recent Refs.~\cite{Dornheim_MRE_2023,Dornheim_PTR_2023}; here, we restrict ourselves to a brief discussion of the key relations that are most relevant to the present work.

First, we consider the frequency moments of the DSF,
\begin{eqnarray}\label{eq:integral}
    \braket{\omega^\alpha} = \int_{-\infty}^\infty \textnormal{d}\omega\ S(\mathbf{q},\omega)\ \omega^\alpha\ ,
\end{eqnarray}
which are related to the derivatives of the ITCF around $\tau=0$~\cite{Dornheim_MRE_2023},
\begin{eqnarray}\label{eq:derivative}
    \braket{\omega^\alpha} = \left. \left(-1\right)^\alpha \frac{\partial^\alpha}{\partial\tau^\alpha} F(\mathbf{q},\tau)\right|_{\tau=0}\, 
\end{eqnarray}
for an integer $\alpha\geq0$. The first moment is defined by the universal f-sum rule~\cite{quantum_theory,Dornheim_SciRep_2024}
\begin{eqnarray}\label{eq:fsum}
    \braket{\omega^1} = -\left.  \frac{\partial}{\partial\tau} F(\mathbf{q},\tau)\right|_{\tau=0} = \frac{{q}^2}{2}\ ,
\end{eqnarray}
which directly explains the increasingly steep slope of the ITCF for decreasing wavelengths $\lambda=2\pi/q$.
We note that Eq.~(\ref{eq:fsum}) has recently been proposed as a model-free means of normalizing the experimentally measured XRTS spectra~\cite{Dornheim_SciRep_2024}.

A second important concept is given by the relative $\tau$-decay measure
\begin{eqnarray} \label{eq:tau_decay}
    \Delta F_\tau(\mathbf{q}) = \frac{F(\mathbf{q},0)-F(\mathbf{q},\tau)}{F(\mathbf{q},0)}\ ,
\end{eqnarray}
which is very sensitive to the particular distribution of spectral weight in the DSF.
Specifically, from the exponential kernel $e^{-\tau\omega}$ in the Laplace transform [Eq.~(\ref{eq:Laplace})], we directly see that spectral weight at large frequencies is associated with a steeper $\tau$-decay; conversely, an exchange--correlation induced red-shift of spectral weight such as the roton-type feature in the strongly coupled electron liquid manifests as a reduced $\tau$-decay and thus as a pronounced feature in Eq.~(\ref{eq:tau_decay})~\cite{Dornheim_MRE_2023}.

Let us finally consider the long wavelength limit, where the DSF is given by a nascent delta peak at the plasma frequency $\omega_\textnormal{p}=\sqrt{3/r_s^3}$,
\begin{eqnarray}\label{eq:DSF_plasmon}
    S_\textnormal{pl}(q\to0,\omega) = A \left\{
\delta(\omega-\omega_\textnormal{pl}) + e^{-\beta\omega_\textnormal{pl}}\delta(\omega+\omega_\textnormal{pl})
    \right\}\ .
\end{eqnarray}
It is straightforward to translate Eq.~(\ref{eq:DSF_plasmon}) into a corresponding $\tau$-decay, leading to
\begin{eqnarray}\label{eq:imaginary_plasmon}
    \Delta F_\textnormal{pl}(q\to0,\tau) = 1 - \frac{e^{-\tau\omega_\textnormal{pl}} + e^{-(\beta-\tau)\omega_\textnormal{pl}} }{1 + e^{-\beta\omega_\textnormal{pl}}}\ ,
\end{eqnarray}
and to the special case of
\begin{eqnarray}\label{eq:imaginary_plasmon_beta_half}
    \Delta F_\textnormal{pl}(q\to0,\beta/2) = 1 - \frac{2}{e^{\beta\omega_\textnormal{pl}/2} + e^{-\beta\omega_\textnormal{pl}/2} }\ .
\end{eqnarray}
The large wavelength limit of the static structure factor $S(\mathbf{q})=F(\mathbf{q},0)$ is given by~\cite{kugler_bounds,review}
\begin{eqnarray}\label{eq:SSF_small_q}
\lim_{q\to0}    S(q) = \frac{q^2}{2\omega_\textnormal{pl}} \textnormal{coth}\left(
\frac{\beta\omega_\textnormal{pl}}{2}
    \right)\ ,
\end{eqnarray}
which follows from the perfect screening in the UEG.
Assuming a delta-like plasmon excitation, Eq.~(\ref{eq:SSF_small_q}) can be generalized to
\begin{eqnarray}\label{eq:Tolias}
    \lim_{q\to0} F(q,\tau) = \frac{q^2}{2\omega_\textnormal{pl}} \frac{\textnormal{cosh}\left[\left(\tau-\frac{\beta}{2}\right)\omega_\textnormal{pl}\right]}{\textnormal{sinh}\left[\frac{\beta\omega_\textnormal{pl}}{2}\right]}\ ;
\end{eqnarray}
see Appendix~\ref{sec:long_wavelength_derivation} for a detailed derivation.

\subsection{Analytic continuation\label{sec:anal_cont_theory}}

For many years, analytic continuation of ITCF data has challenged the scientific community~\cite{Otsuki_PRE_2017,Otsuki_JPSJ_2020, JARRELL1996133, BurnierRothkopf2013bayesianreconstruction, Boninsegni_maximum_entropy, Silver_PRB_1990, Mishchenko_PRB_2000,Vitali_PRB_2010,Filinov_PRA_2012,Filinov_PRA_2016,Ferre_PRB_2016, shu2015stochastic, Dornheim_SciRep_2022, hansen2019extraction, tripolt2019comparison, Fournier_PRL_2020,Yoon_PRB_2018, tkachenko_book,Tkachenko_CPP_2018,Vorberger_PRL_2012, Filinov_PRB_2023, Loon_PRB_2016,Goulko_PRB_2017, tripolt2019comparison, huang2023acflow, KAUFMANN2023ana_cont}. 
However, for the analytic continuation of UEG ITCF data, recent investigation have produced model-free (\textit{i.e.}, unparameterized) dynamic structure factors \cite{chuna2025dualformulationmaximumentropy, chuna2025estimatesdynamicstructurefactor, pylit, Filinov_PRB_2023}. Each investigation was conducted with a different method and, although all-encompassing quantitative comparisons have not been made, there is good qualitative agreement with each other and the earlier discussed parameterized stochastic sampling approach~\cite{dynamic_folgepaper, Dornheim_PRE_2020}. Furthermore, based on equations \eqref{eq:integral} and \eqref{eq:derivative}, we see that matching the ITCF data well enforces the sum rules that are also present in the data; this is discussed more in Chuna et al.~\cite{chuna2025estimatesdynamicstructurefactor}. Thus, being model-free does not imply that sum-rules (\textit{e.g.}, f-sum rule \eqref{eq:fsum}) are not satisfied. In summary, reliable estimates of the DSF can be produced. 

Here, we use the PyLIT code \cite{pylit} to analytically continue the ITCF data used in this investigation. At large $q$, where the DSF occupies large $\omega$ values the ITCF decays rapidly compared to its sampling; this leaves only a few data points that are not numerically equivalent to zero. As such typical formulations using uniform $\omega$ grid resolution break down. PyLIT's kernel based approach is naturally suited to overcome this difficulty. 

We briefly discuss the theory behind PyLIT's implementation. The PyLIT code represents the DSF as a linear combination of kernels with analytic Laplace transforms. 
\begin{align}
    S_\mathbf{\alpha}(\omega) = \sum_{j=1}^{m} \mathbf{\alpha}_j K_j(\omega).
\end{align}
The Laplace transform is a linear operator so the inverse Laplace transform is reformulated as fitting a linear combination of Laplace transformed kernels $R$ to the ITCF signal $\mathbf{F}$. The reformulated problem is expressed as
\begin{align}\label{eq:min_problem}
        \underset{\mathbf{\alpha}}{\min} \, \, \frac{1}{2} \Vert \mathbf{F} - R \mathbf{\alpha} \Vert_2^2 + \lambda r(\cdot)\ .
\end{align}
This reformulated problem if preferred because the kernels can be selected intelligently, with a non-uniform spacing, according to a default model (\textit{i.e.}, Bayesian prior). This leads to fewer unknowns than the typical formulations (\textit{i.e.} reduces the search space). While the search space has been reduced the problem is still ill-conditioned. Thus, a regularization term $r(\cdot)$ has been introduced in \eqref{eq:min_problem}.

In this work, we only regularize using the Wasserstein distance \cite{yang2023wasserstein}
\begin{subequations}\label{eq:regularization}
\begin{align}\label{eq:Wasserstein}
    r(\cdot) = \frac{1}{2}\left\|\operatorname{CDF}[S_\mathbf{\alpha}(\omega)-D(\omega)]\right\|_2^2,
\end{align}
where the CDF is defined as
\begin{equation}\label{eq:CDF}
    \mathrm{CDF}[S_\mathbf{\alpha}](\omega) = \int_{-\infty}^\omega S_\mathbf{\alpha}(\omega') \, \text{d} \omega'. 
%    \mathrm{CDF}[S_\ba]: L^1(\mathbb{R}) \rightarrow L^1_{\mathrm{loc}}(\mathbb{R}), \quad S_\ba \mapsto \int_{-\infty}^x S_\ba(y) \dy,
\end{equation}
\end{subequations}
The choice of the regularization weight is handled by a Bayesian posterior weighting scheme~\cite{gull1989MEMBayesianWeighting}. The interested reader may explore the PyLIT publication~\cite{pylit}, where other regularizations and regularization weight selection procedures are investigated.

\begin{figure*}\centering
\includegraphics[width=0.32\textwidth]{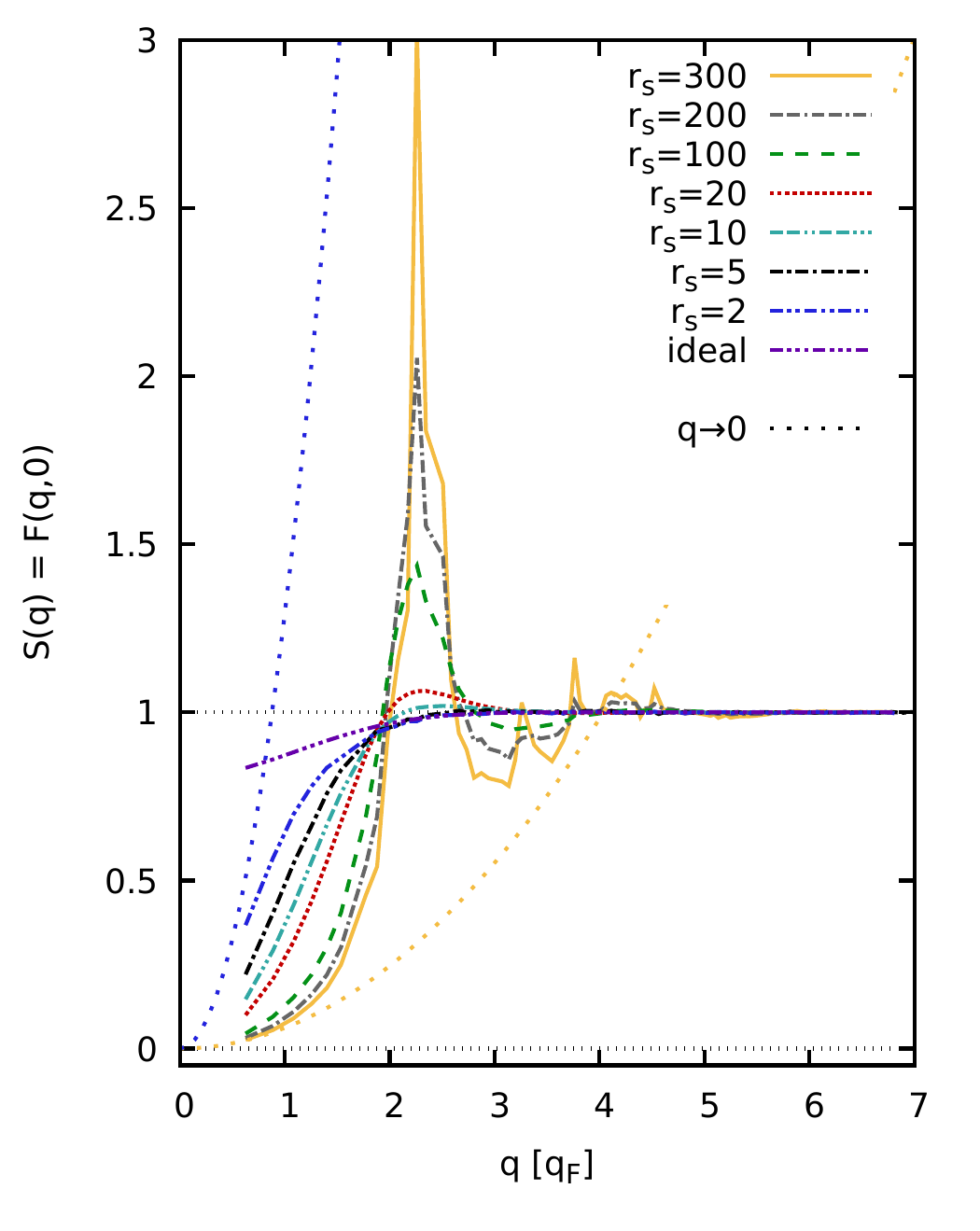}%
\includegraphics[width=0.32\textwidth]{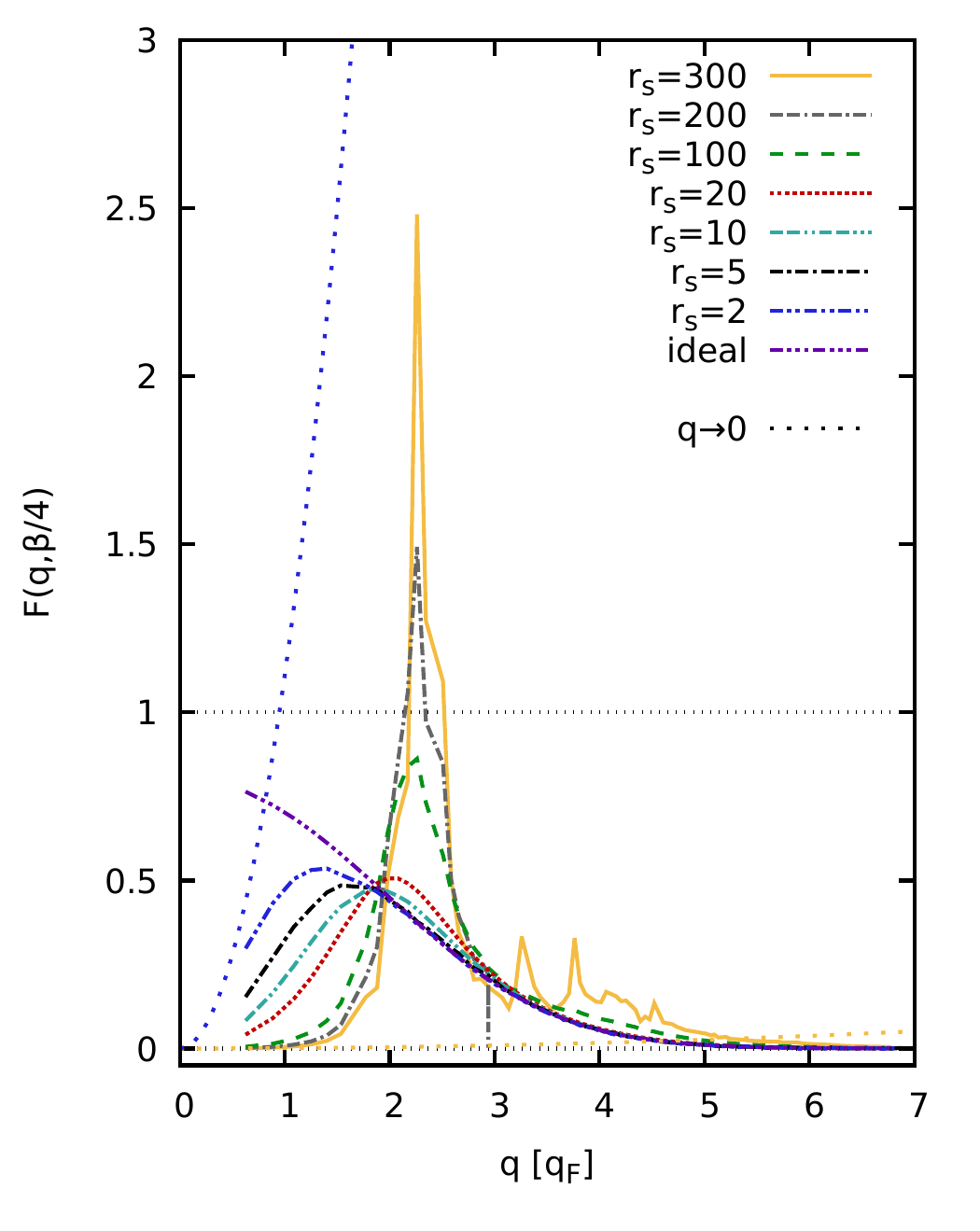}%
\includegraphics[width=0.32\textwidth]{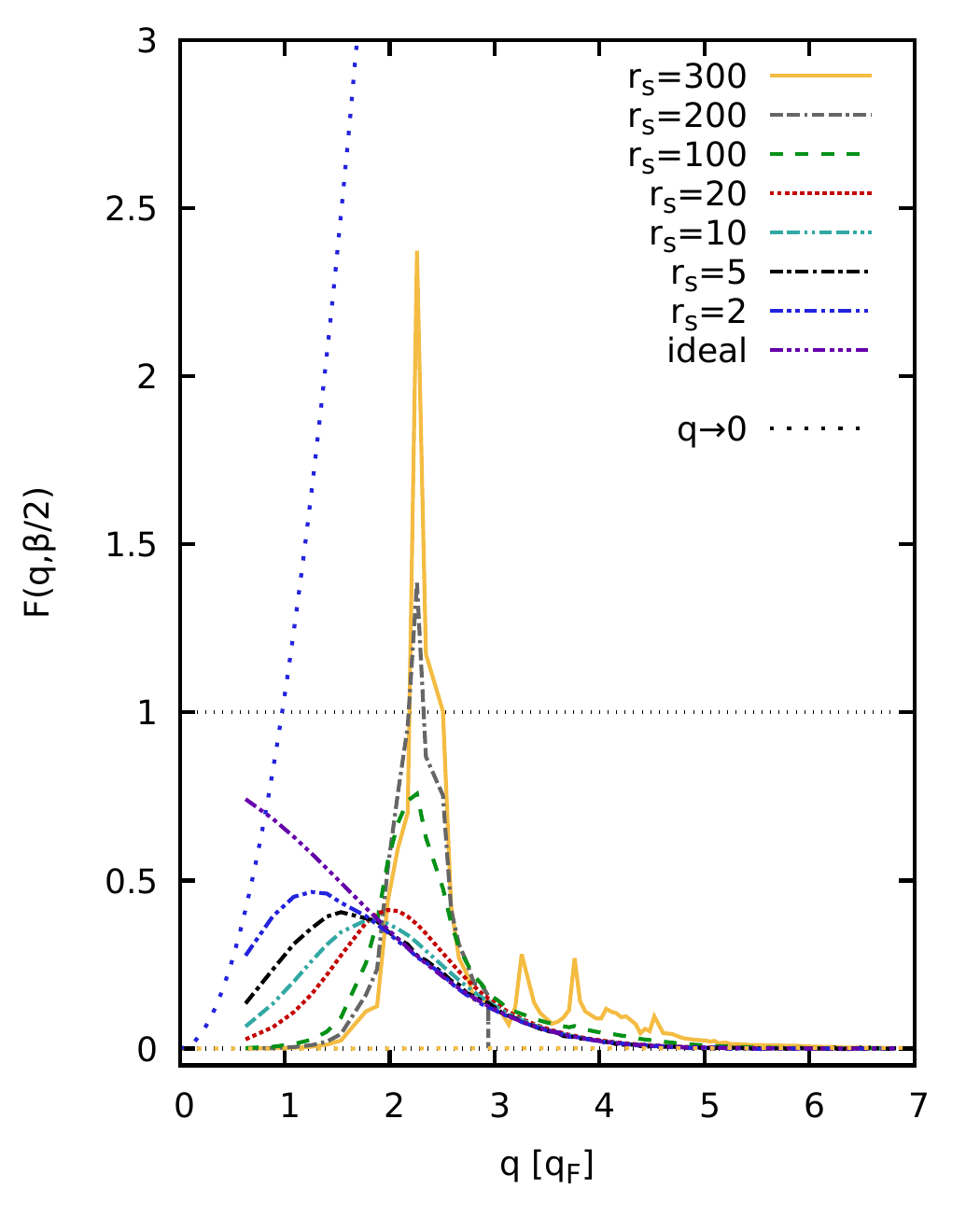}
\caption{\emph{Ab initio} PIMC results for the static structure factor $S(\mathbf{q})=F(\mathbf{q},0)$ [left panel], the ITCF at a quarter of the inverse temperature $F(\mathbf{q},\beta/4)$ [central panel], and the thermal structure factor $S_{\beta/2}(\mathbf{q})=F(\mathbf{q},\beta/2)$ [right panel] for $N=34$ unpolarized electrons at $\Theta=1$ and various coupling parameters. The blue and yellow dotted lines show the large wavelength asymptotic of the ITCF, Eq.~(\ref{eq:Tolias}).}
\label{fig:SSF} 
\end{figure*}

\section{Results\label{sec:results}}

All PIMC results presented in this work have been obtained using the open-source \texttt{ISHTAR} code~\cite{ISHTAR} using the extended canonical ensemble sampling scheme introduced in Ref.~\cite{Dornheim_PRB_nk_2021}, and are freely available in an online repository~\cite{repo}.

\subsection{Imaginary-time perspective\label{sec:imaginary}}

In Fig.~\ref{fig:SSF}, we show our PIMC results for $N=34$ unpolarized electrons at $\Theta=1$ for a broad range of densities, $2\leq r_s\leq300$. The left panel corresponds to the static structure factor $S(\mathbf{q})=F(\mathbf{q},0)$ and exhibits a pronounced dependence on the density. In the long wavelength limit of $q\to0$, all UEG curves exhibit a parabolic shape that is described by Eq.~(\ref{eq:SSF_small_q}), see the dotted blue and yellow curves for $r_s=2$ and $r_s=300$, which directly follows from the perfect screening in the UEG~\cite{kugler_bounds}. For completeness, we have included results for an ideal Fermi gas at $\Theta=1$ (which essentially corresponds to $r_s\to0$) as the dash-triple-dotted purple line. In the noninteracting case, $S(\mathbf{q})$ attains a finite value $\in(0,1)$ in the limit of $q\to0$ due to exchange effects. Conversely, all curves converge towards unity in the short wavelength limit of large $q$, which is a consequence of perfect correlations of each electron with itself. We note that the particular form of the convergence is related to the on-top pair correlation function $g(0)$, see, e.g., Ref.~\cite{Hunger_PRE_2021}. Arguably, the most interesting regime is given by intermediate wavenumbers $q\sim2q_\textnormal{F}$, where the associated wavelength $\lambda=2\pi/q$ is comparable to the average interparticle distance~\cite{Dornheim_Nature_2022}. Here, $S(\mathbf{q})$ exhibits a peak for $r_s\gtrsim10$, which becomes increasingly pronounced for $r_s\gtrsim 100$. This is the hallmark of the strongly coupled electron liquid regime and indicates the presence of strong interparticle correlations. 
For $r_s\gtrsim100$, the peak is followed by a minimum and, in the case of $r_s=300$, by a second peak and a shallow, but clearly visible, second minimum.

At this point, we feel that a note on the finite system size in our PIMC simulations is pertinent. For $r_s\lesssim100$, all depicted curves are very smooth and it is well known that finite-size effects in $q$-resolved properties such as the ITCF and, hence, $S(\mathbf{q})$ are very small~\cite{dornheim_prl,Chiesa_PRL_2006,Holzmann_PRB_2016,Drummond_PRB_2008,Dornheim_JCP_2021,dornheim2025applicationsphericallyaveragedpair}. Instead, the main effect of the simulated number of particles is given by the discrete $q$-grid that follows from momentum quantization in the finite simulation cell. For $r_s\gtrsim100$, we witness the formation of correlations on medium length scales, which changes this situation. Specifically, PIMC results are affected by commensurability effects, as the finite length of the main cell shapes the spatial orientation of the electrons; this is the origin of the small spikes in $S(\mathbf{q})$, which are particularly well visible for $r_s=300$. Indeed, it is well known that at least $N\sim10^3-10^4$ electrons are required to capture the correct symmetry of the Wigner crystal~\cite{Clark_PRL_2009}, both of which, however, is beyond the scope of the present work. Therefore, we here limit ourselves to the electron liquid regime of $r_s\leq300$, where we can still draw meaningful conclusions from the present simulations with $N=34$.

Let us next come to the topic at hand, which is the detailed analysis of the ITCF $F(\mathbf{q},\tau)$. In the center and the right panels of Fig.~\ref{fig:SSF}, we show the $q$-dependence of $F(\mathbf{q},\tau)$ for $\tau=\beta/4$ and $\tau=\beta/2$, respectively. The latter is also known as the \emph{thermal structure factor} in the literature~\cite{Dornheim_MRE_2023}. Overall, we find very similar qualitative trends for the two values of $\tau$, which, however, substantially differ from the case of $\tau=0$ shown in the left panel of the same figure. In particular, $F(\mathbf{q},\tau)$ converges towards zero in the short wavelength limit for any finite $\tau$, as self-correlations, too, decay along the Gaussian imaginary-time diffusion process~\cite{Dornheim_MRE_2023,Dornheim_PTR_2023}. The increasing decay of correlations also affects the intermediate $q$-regime, where the peak(s) is (are) now located on top of the ideal results. The long wavelength regime is still predominantly shaped by the perfect screening sum rule, which can be generalized to Eq.~(\ref{eq:Tolias}) for the entire $\tau$-dependence of the ITCF at small $q$, see the dotted blue and yellow lines. It is pointed out that, for an ideal Fermi gas, the long wavelength limit of the ITCF does not depend on the imaginary time~\cite{Tolias_CtPP2025}, which is confirmed by extrapolating the PIMC results of Fig.~\ref{fig:SSF}. We also note that the observed qualitative similarity of $F(\mathbf{q},\beta/2)$ and $F(\mathbf{q},\beta/4)$ might be of potential value for the interpretation of XRTS experiments, where systematic uncertainties are known to increase with increasing $\tau$~\cite{Dornheim_T_2022}.

\begin{figure}\centering
\includegraphics[width=0.449\textwidth]{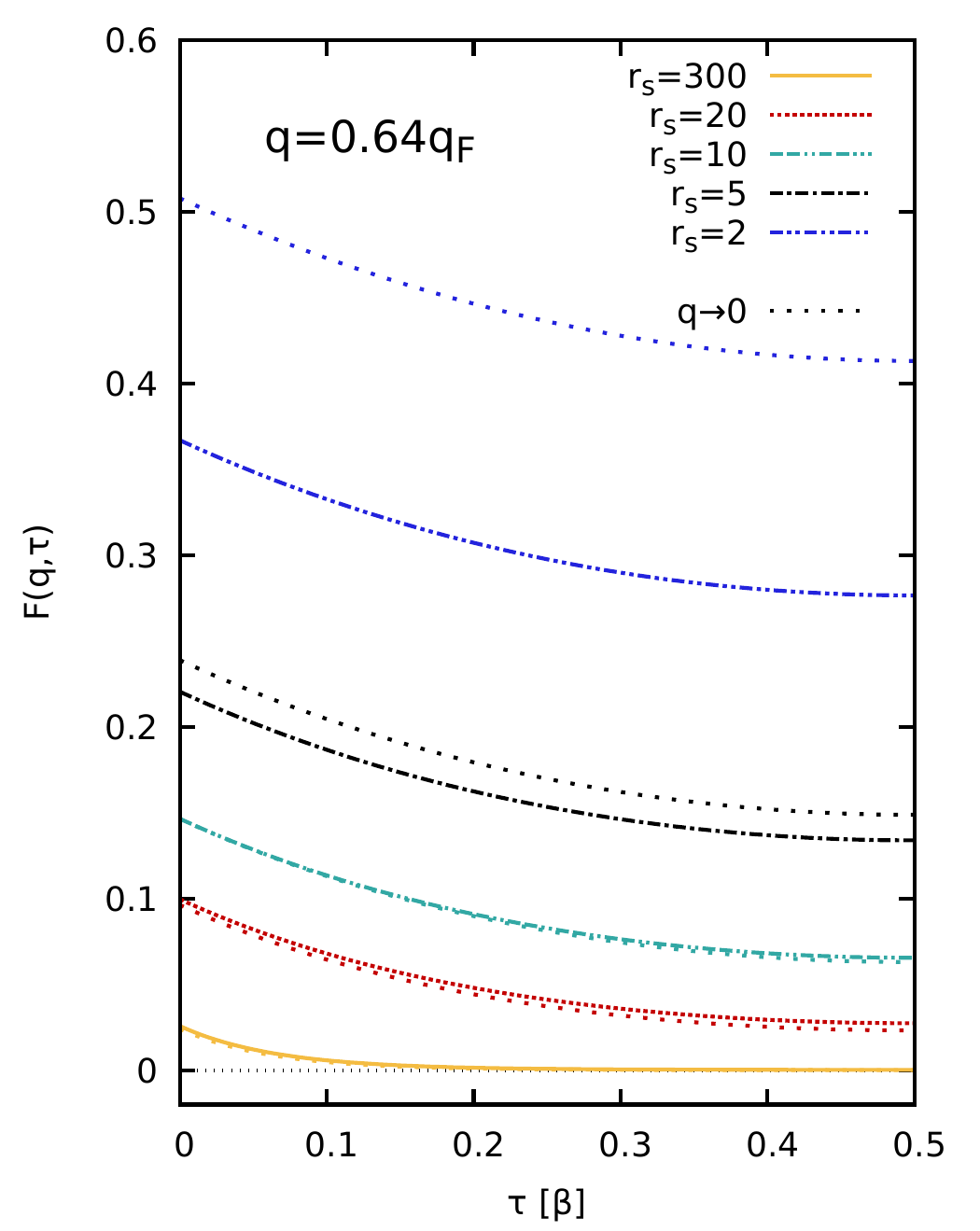}
\caption{\label{fig:Large_Wavelength} \emph{Ab initio} PIMC results for the ITCF $F(\mathbf{q},\tau)$ at the smallest wavenumber $q=2\pi/L$ ($q=0.64q_\textnormal{F}$), for $N=34$ unpolarized electrons at $\Theta=1$ and various coupling parameters. Dotted curves show the large wavelength limit of Eq.~(\ref{eq:Tolias}).}
\end{figure} 

In Fig.~\ref{fig:Large_Wavelength}, we gather our PIMC results for the $\tau$-dependence of the ITCF at the smallest possible wavenumber $q=2\pi/L$ for $N=34$, where the dotted lines correspond to the long wavelength limit, Eq.~(\ref{eq:Tolias}). We note that the ITCF obeys the detailed balance relation $F(\mathbf{q},\beta-\tau)=F(\mathbf{q},\tau)$~\cite{Dornheim_Nature_2022}, so we restrict ourselves to the non-redundant half-range of $\tau\in[0,\beta/2]$ here. Evidently, Eq.~(\ref{eq:Tolias}) nicely captures the $\tau$-dependence for all densities. In addition, it is in very good agreement for the full data for $r_s\gtrsim10$, but starts to deviate at higher densities. This is a direct consequence of the waning agreement in $S(\mathbf{q})=F(\mathbf{q},0)$ in this regime and has to do with the characteristic length scale of the Coulomb interaction. Specifically, we have smaller simulation cells for higher densities at $N=34=\textnormal{const}$, which shifts the onset of the plasmon limit to smaller values of $q/q_\textnormal{F}$.

\begin{figure}\centering
\includegraphics[width=0.49\textwidth]{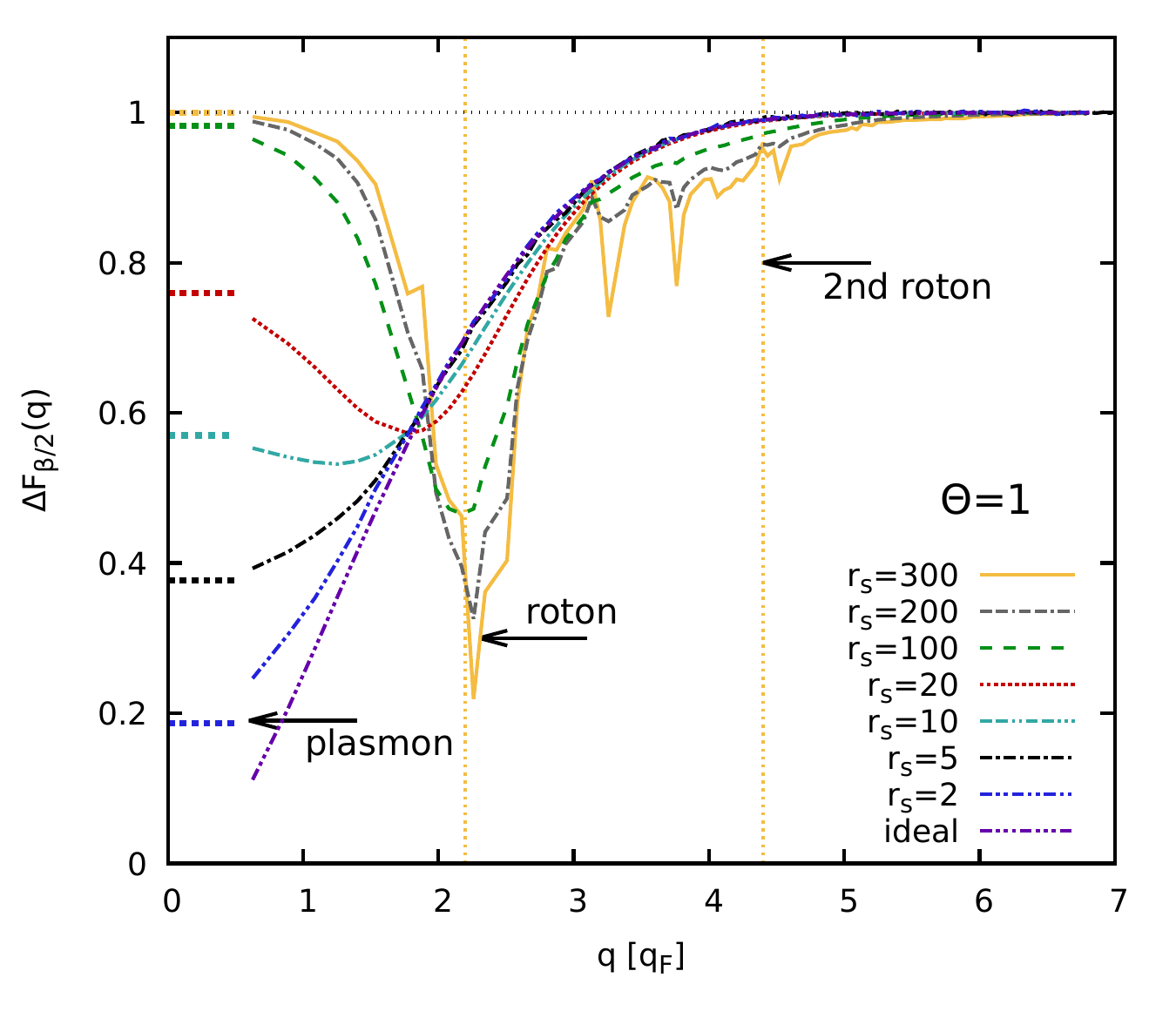}\\\vspace*{-1.05cm}\includegraphics[width=0.49\textwidth]{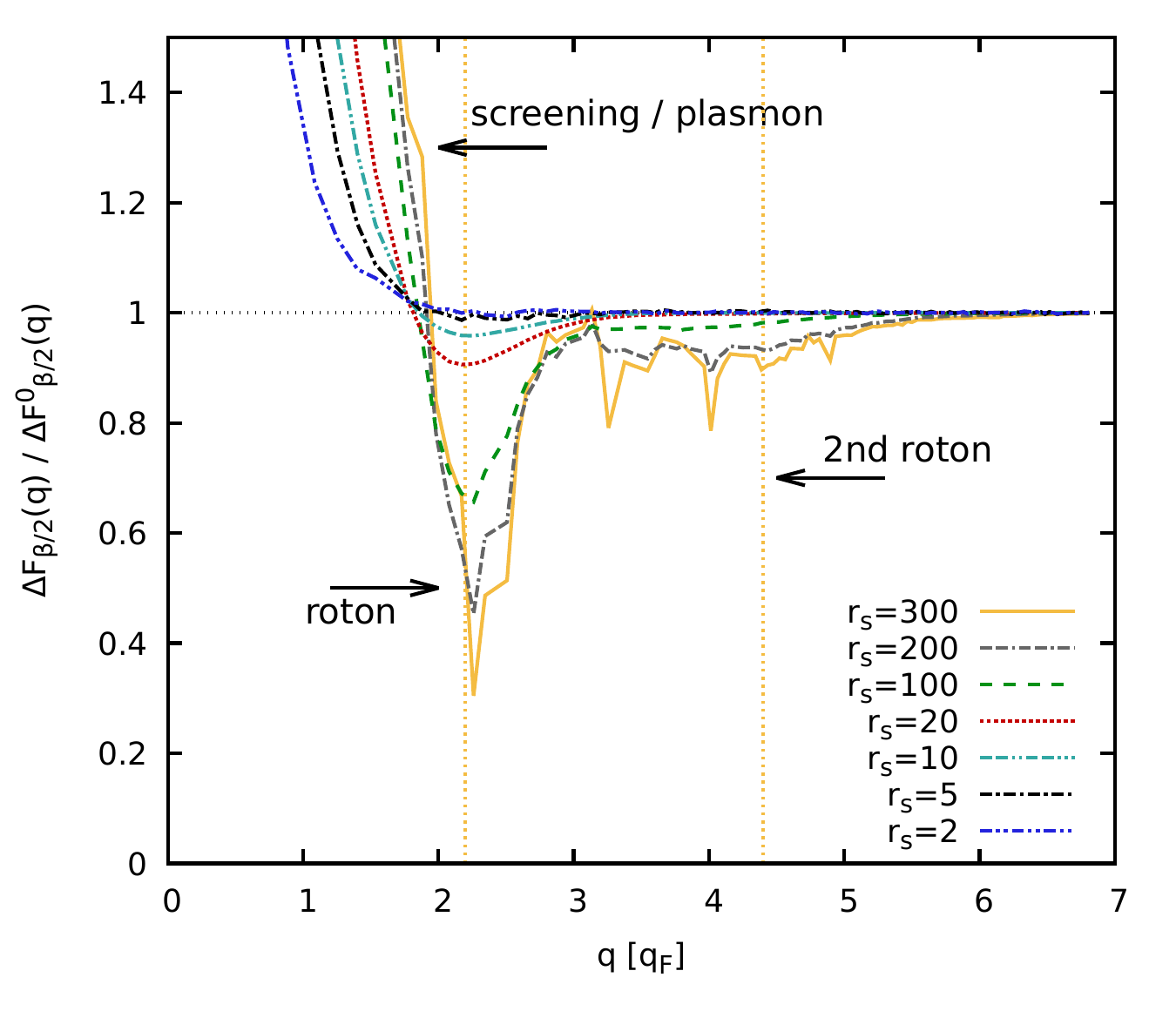}
\caption{\label{fig:imaginary_dispersion} Top: \emph{Ab initio} PIMC results for the relative $\tau$-decay $\Delta F_{\beta/2}(\mathbf{q})$ [Eq.~(\ref{eq:tau_decay})] of the ITCF at $\Theta=1$ and various coupling parameters. The horizontal bars indicate the plasmonic limit, Eq.~(\ref{eq:imaginary_plasmon_beta_half}), and the vertical yellow lines the $q$-regime of the first and second roton. Bottom: The relative $\tau$-decay $\Delta F_{\beta/2}(\mathbf{q})$ re-normalized by the  corresponding relative $\tau$-decay of the ideal Fermi gas $\Delta F^0_{\beta/2}(\mathbf{q})$.
}
\end{figure} 

\begin{figure*}\centering
\includegraphics[width=0.329\textwidth]{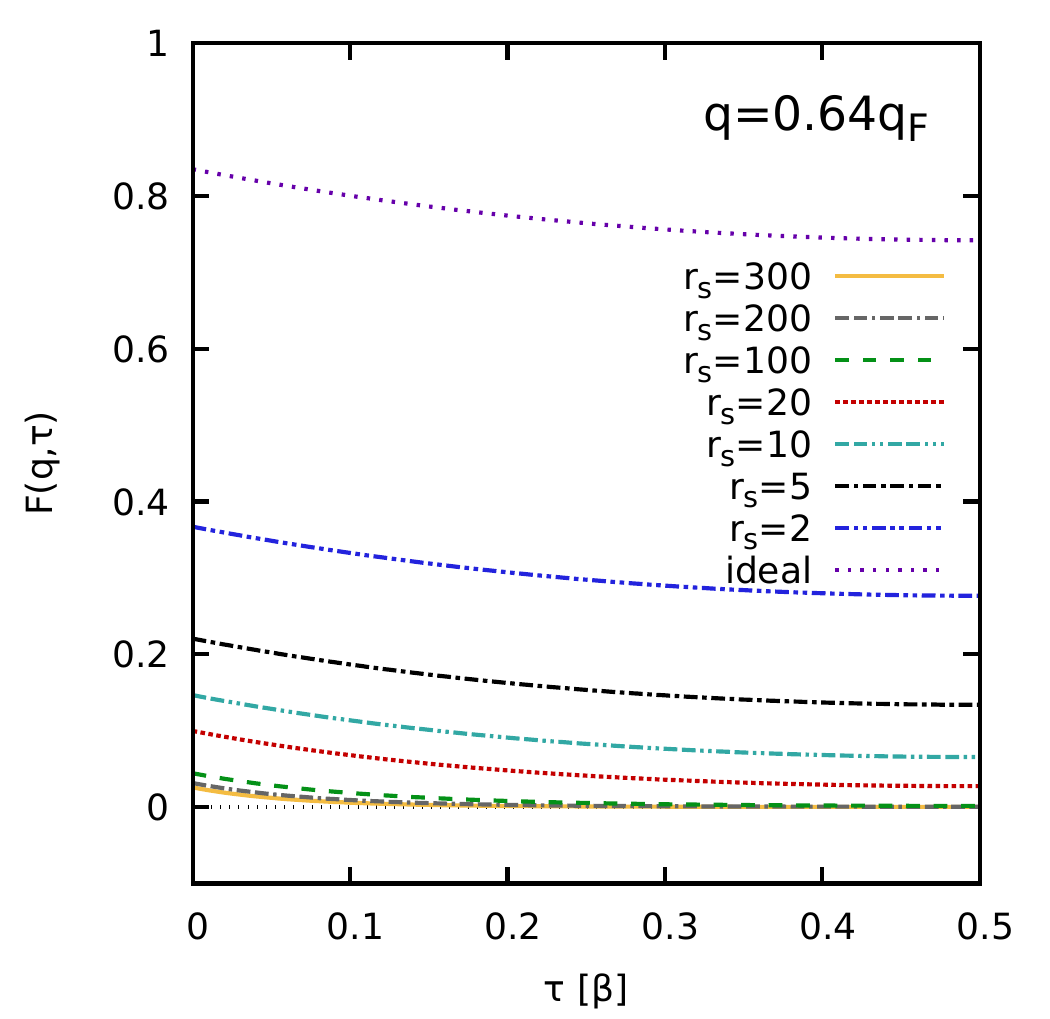}\includegraphics[width=0.329\textwidth]{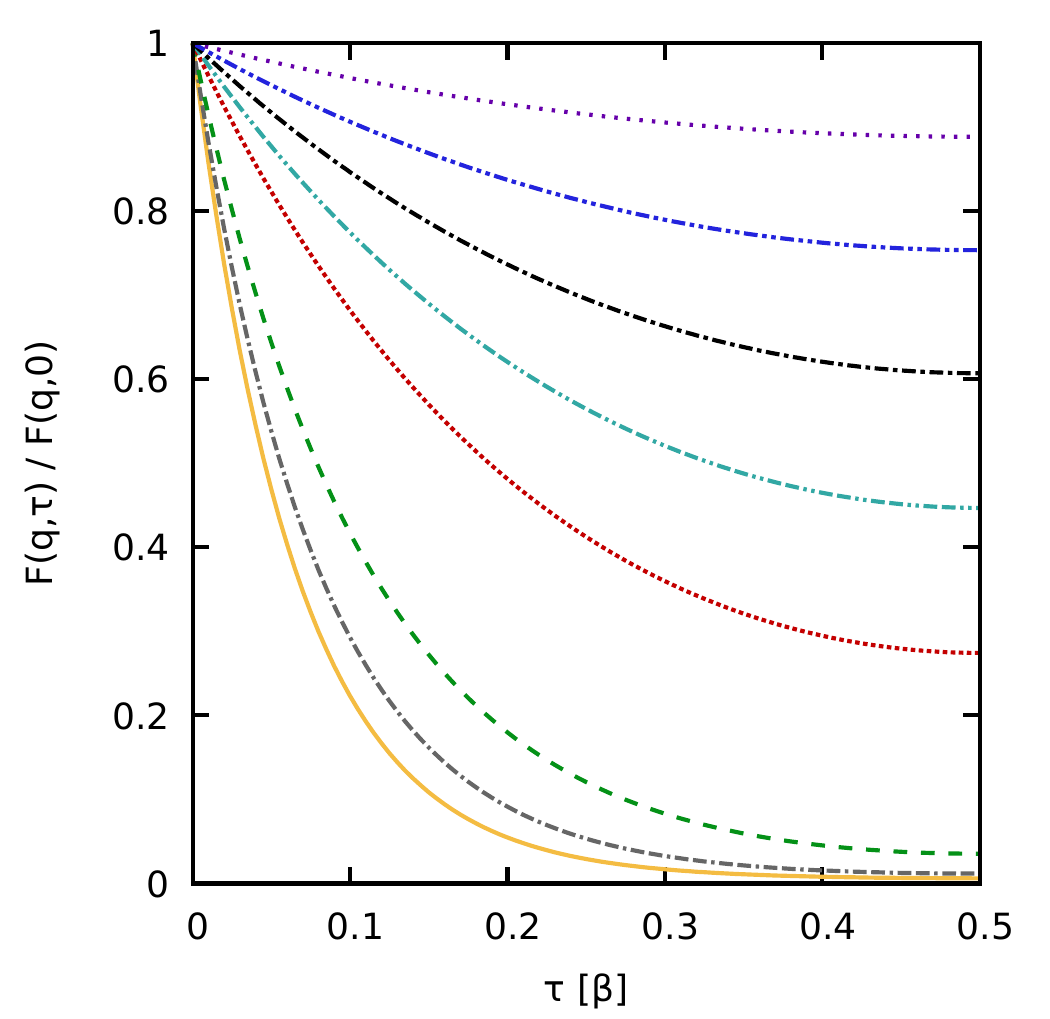}\includegraphics[width=0.329\textwidth]{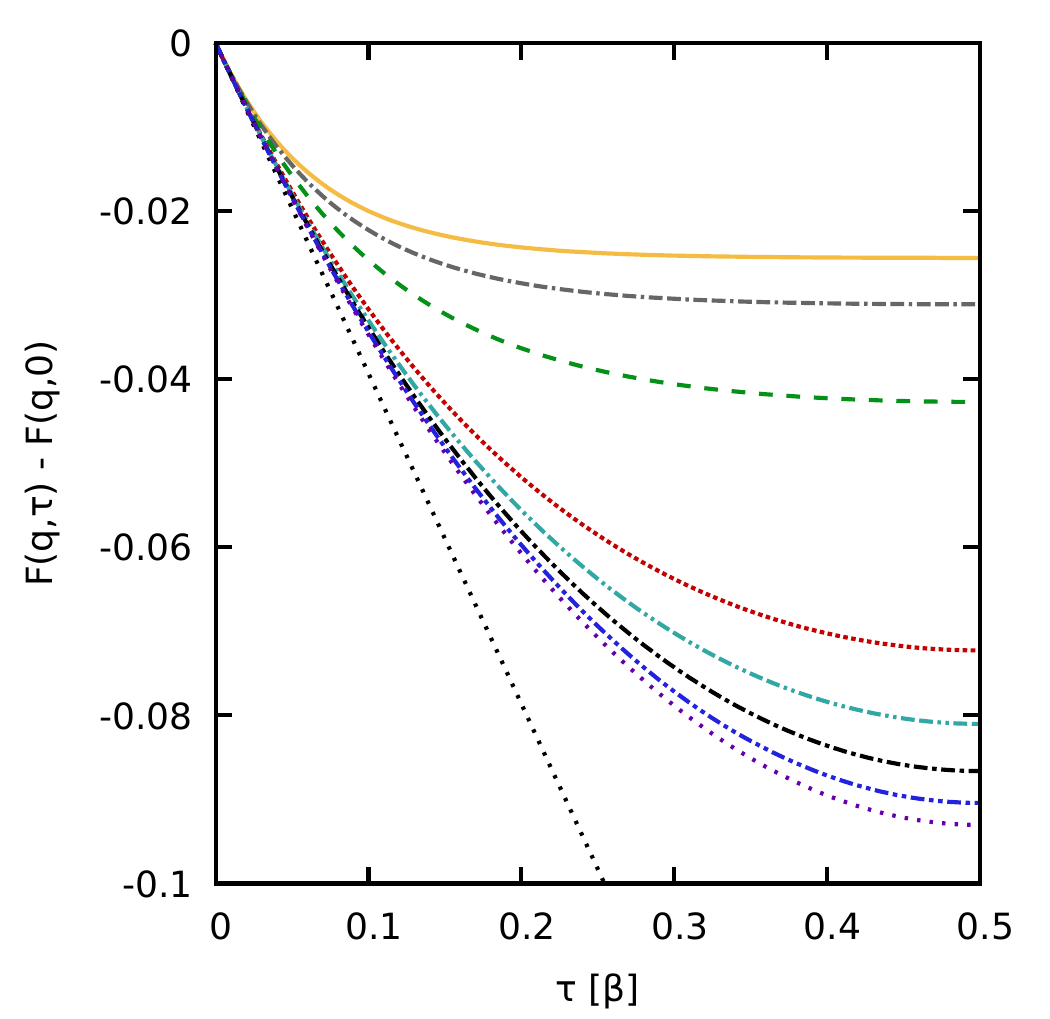}\\
\includegraphics[width=0.329\textwidth]{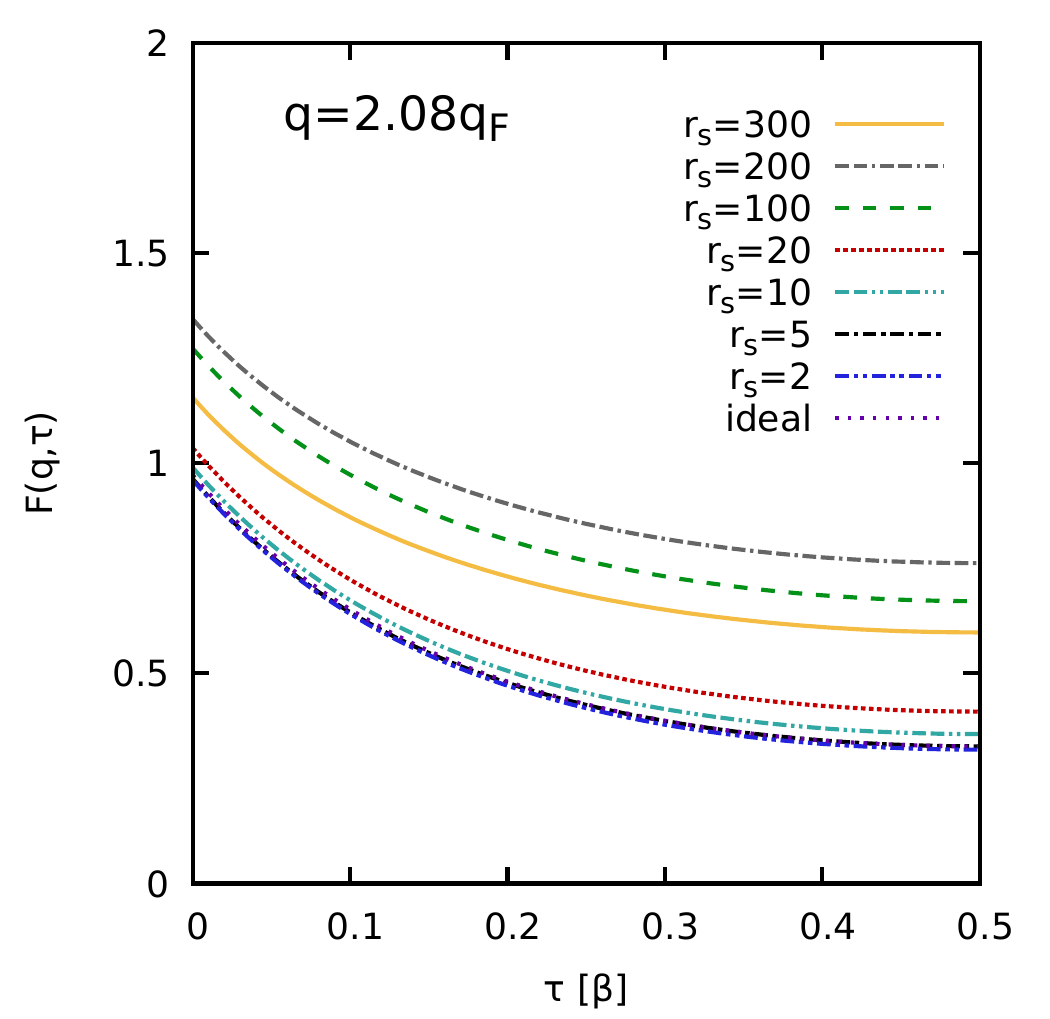}\includegraphics[width=0.329\textwidth]{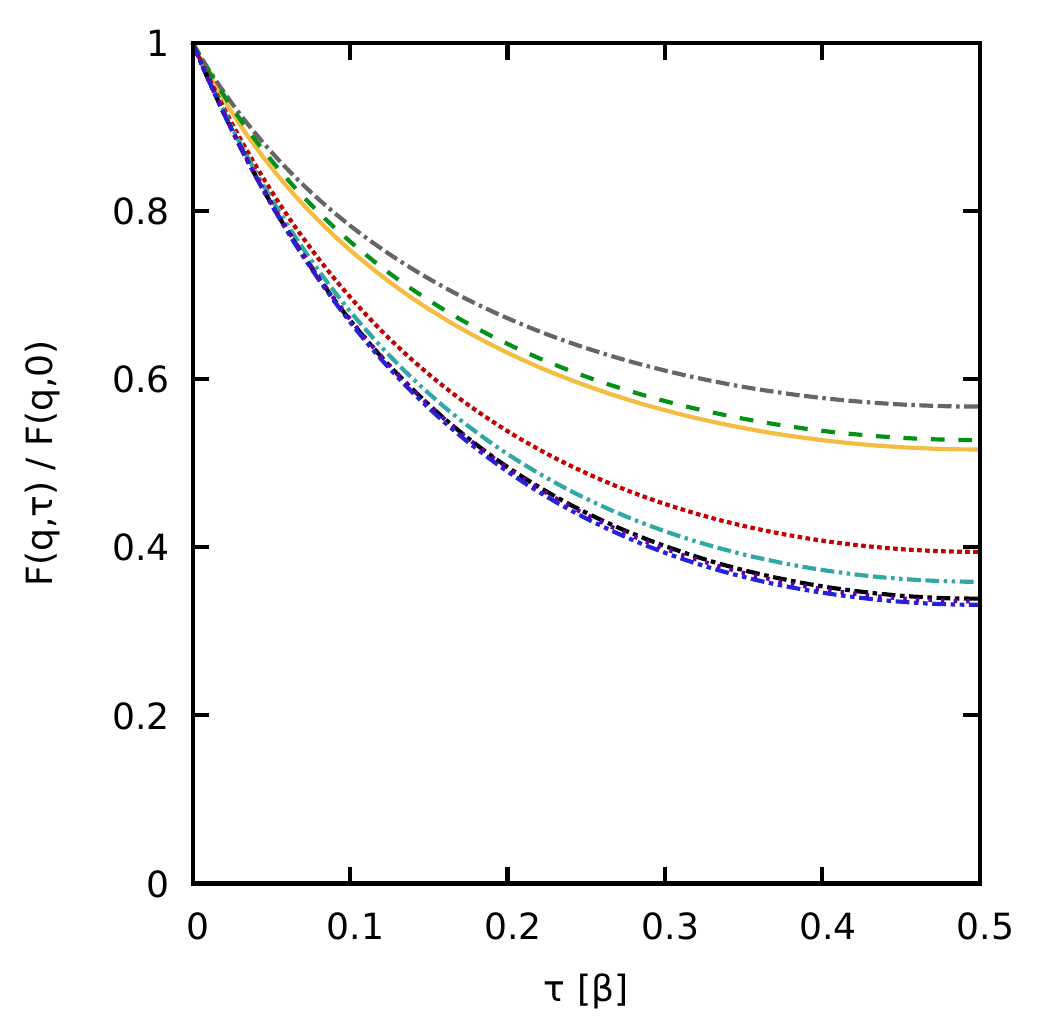}\includegraphics[width=0.329\textwidth]{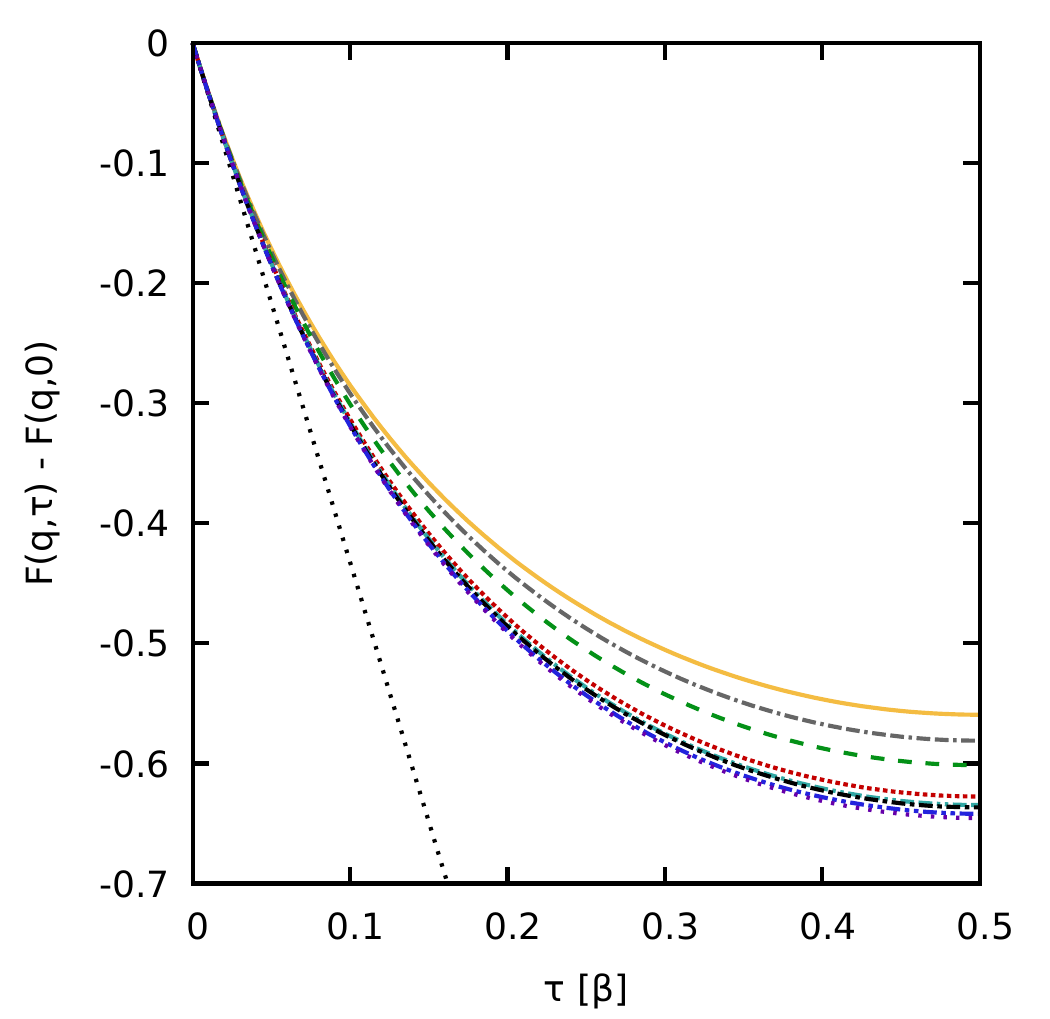}\\
\includegraphics[width=0.329\textwidth]{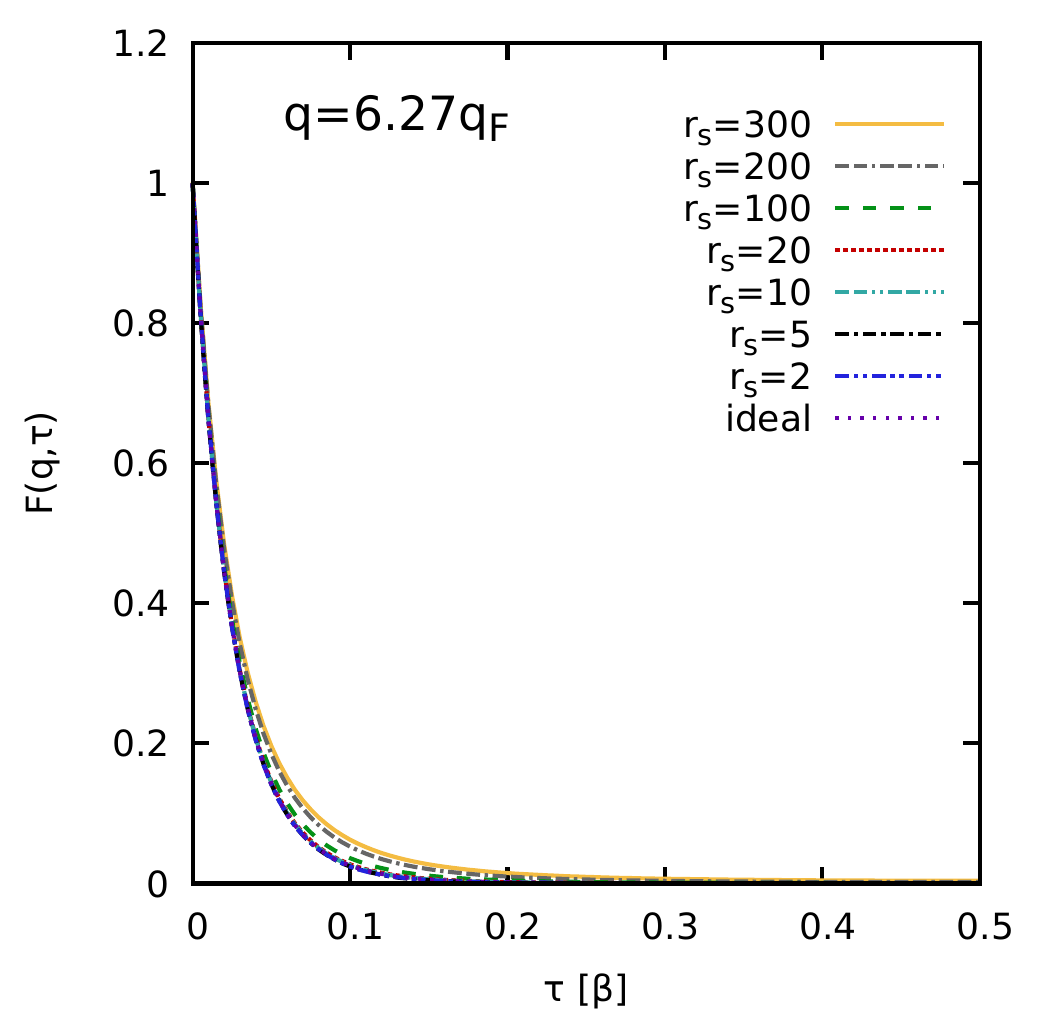}\includegraphics[width=0.329\textwidth]{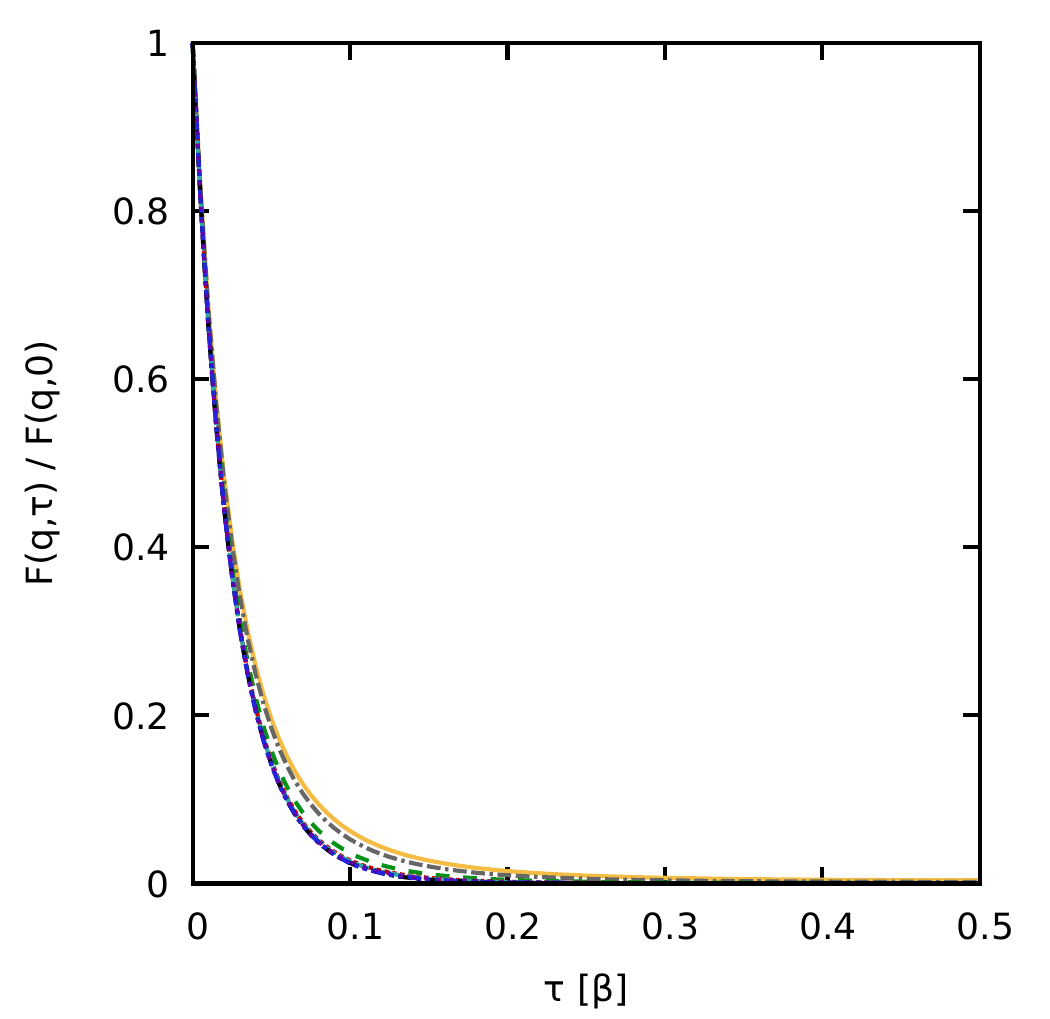}\includegraphics[width=0.329\textwidth]{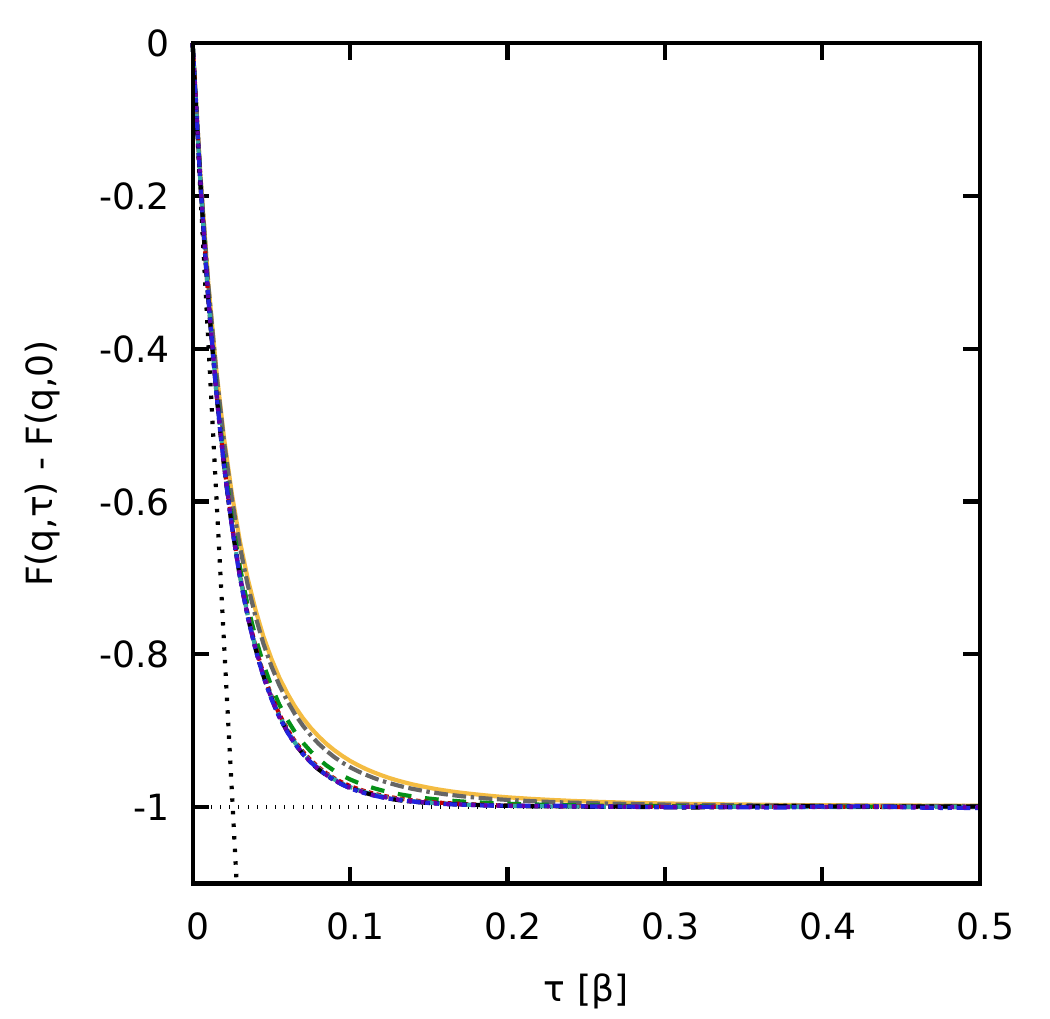}
\caption{\label{fig:ITCF} \emph{Ab initio} PIMC results for the ITCF $F(\mathbf{q},\tau)$ at $\Theta=1$ and various coupling parameters. The top, center, and bottom rows correspond to $q=0.64q_\textnormal{F}$, $q=2.08q_\textnormal{F}$, and $q=6.27q_\textnormal{F}$. The left, center, and right columns show $F(\mathbf{q},\tau)$, $F(\mathbf{q},\tau)/F(\mathbf{q},0)$ and $F(\mathbf{q},\tau)-F(\mathbf{q},0)$, respectively. The dotted grey lines in the latter columns show the universal f-sum rule, Eq.~(\ref{eq:fsum}).}
\end{figure*}

Let us next investigate the relative $\tau$-decay $\Delta F_{\beta/2}(\mathbf{q})$, see~Eq.~(\ref{eq:tau_decay}); the corresponding PIMC results are shown in the top panel of Fig.~\ref{fig:imaginary_dispersion}. \emph{First}, we find that all UEG data sets converge towards the correct plasmonic limit for $q\to0$, see Eq.~(\ref{eq:imaginary_plasmon_beta_half}) and the horizontal dotted bars, which is also consistent with the expansion given in Eq.~(\ref{eq:Tolias}). Interestingly, the plasmonic limit of $\Delta F_{\beta/2}(\mathbf{q})$ is ordered with respect to the density. It is easy to see that $\beta\hbar\omega_\textnormal{pl}\sim\sqrt{r_s}/\Theta$, which leads to the strong coupling asymptotic $\lim_{r_s\to\infty}\Delta F_\textnormal{pl}(\beta/2) = 1$. Conversely, the relative $\tau$-decay vanishes for $q\to0$ in the ideal limit of $r_s\to0$~\cite{Tolias_CtPP2025}. \emph{Second}, it is evident that $\Delta F_{\beta/2}(\mathbf{q})$ converges to unity in the short wavelength limit as $F(\mathbf{q}\to\infty,\beta/2)\to0$, cf.~Fig.~\ref{fig:ITCF}. \emph{Third}, we observe a pronounced roton feature at intermediate wavenumbers, $q\sim2q_\textnormal{F}$, as expected. It should be noted that the corresponding peak in $S(\mathbf{q})=F(\mathbf{q},0)$, which has been shown in Fig.~\ref{fig:SSF}, is not sufficient to verify the roton, since it could potentially have been caused by a simple scaling of the full DSF at all frequencies instead of a down-shift of the spectral weight as it is required. By design, the definition of $\Delta F_\tau(\mathbf{q})$ rules out this possibility due to the normalization by $F(\mathbf{q},0)$. \emph{Fourth}, we find clear signatures of a second roton feature at roughly twice the roton wavenumber for $r_s\gtrsim100$. This can be discerned particularly well in the bottom panel of Fig.~\ref{fig:imaginary_dispersion}, where we have renormalized $\Delta F_{\beta/2}(\mathbf{q})$ by the corresponding results for the ideal Fermi gas. Within this representation, all curves converge to the same single-particle asymptotic for large $q$, whereas the UEG exhibits a reduced relative $\tau$-decay in the vicinity of the first and second roton features. Conversely, there is no screening in the ideal Fermi gas, which leads to diverging curves in the limit of $q\to0$.

\begin{figure}\centering
\includegraphics[width=0.49\textwidth]{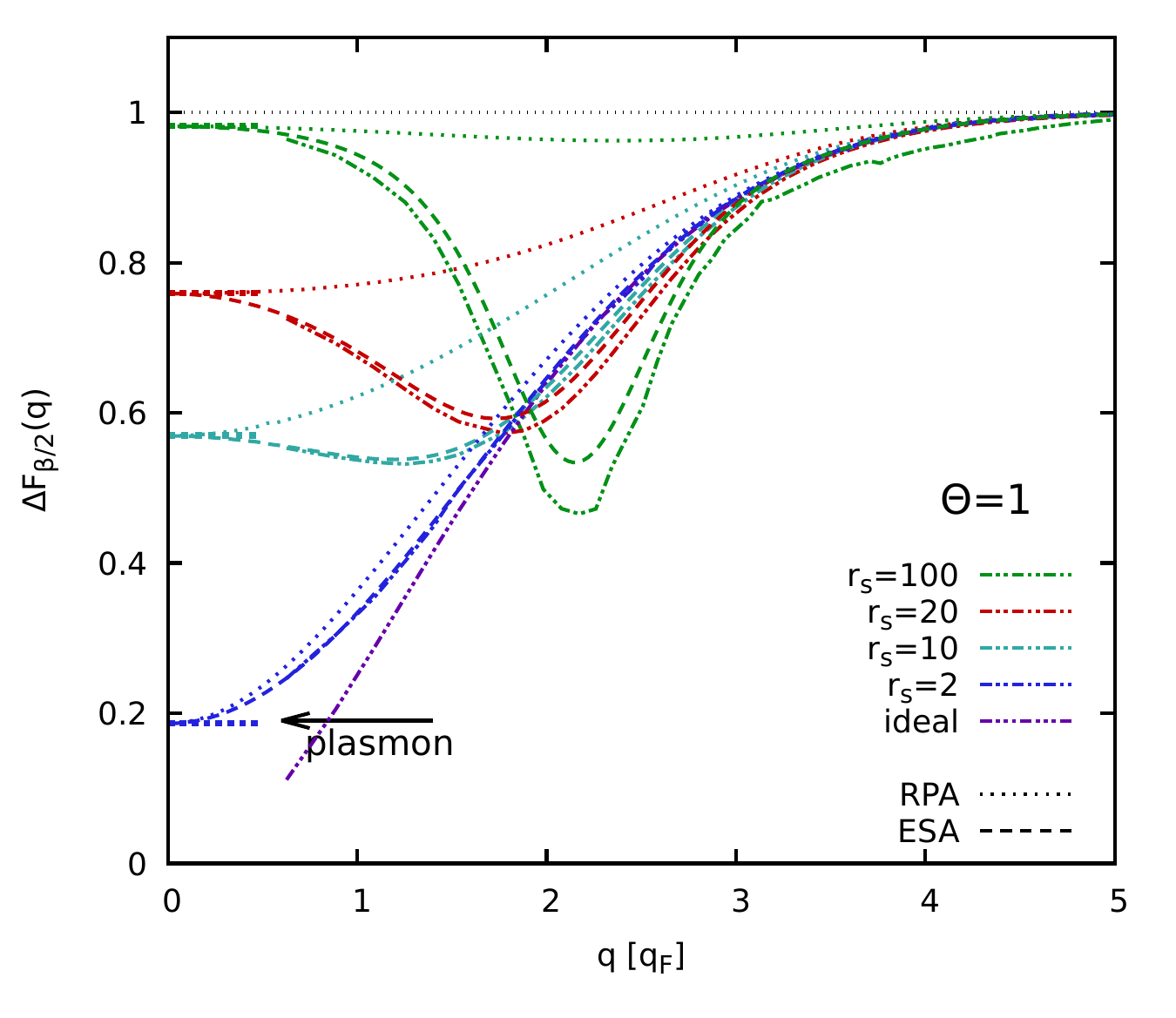}
\caption{\label{fig:theory_imaginary_dispersion} Relative $\tau$-decay $\Delta F_{\beta/2}(\mathbf{q})$ [Eq.~(\ref{eq:tau_decay})] of the ITCF at $\Theta=1$ and various coupling parameters. \emph{Ab initio} PIMC results (solid lines), RPA [$G(\mathbf{q},\omega)\equiv0$] results (dotted lines) and static approximation [$G(\mathbf{q},\omega)\equiv G(\mathbf{q},0)$] results (dashed lines).}
\end{figure} 

Let us postpone the discussion of the physical origin of the second roton and first consider the capability of different approximations to capture these trends. In Fig.~\ref{fig:theory_imaginary_dispersion}, we compare our \emph{ab initio} PIMC results for selected values of $r_s$ to corresponding data sets obtained within the RPA (dotted) and within the \emph{static approximation} (dashed). Evidently, all curves attain the correct plasmon limit, as it is expected.
Interestingly, the RPA is not only incapable to even qualitatively predict the roton feature at $q\sim2q_\textnormal{F}$, see also the extensive discussion in Ref.~\cite{Dornheim_Nature_2022}, but starts to rapidly deviate from the PIMC data for $q>0$. In contrast, the \emph{static approximation} provides the correct qualitative picture even for $r_s=100$. The latter was calculated using the analytical fit for the ESA as given in Ref.~\cite{Dornheim_PRB_ESA_2021}.

\begin{figure}\centering
\includegraphics[width=0.49\textwidth]{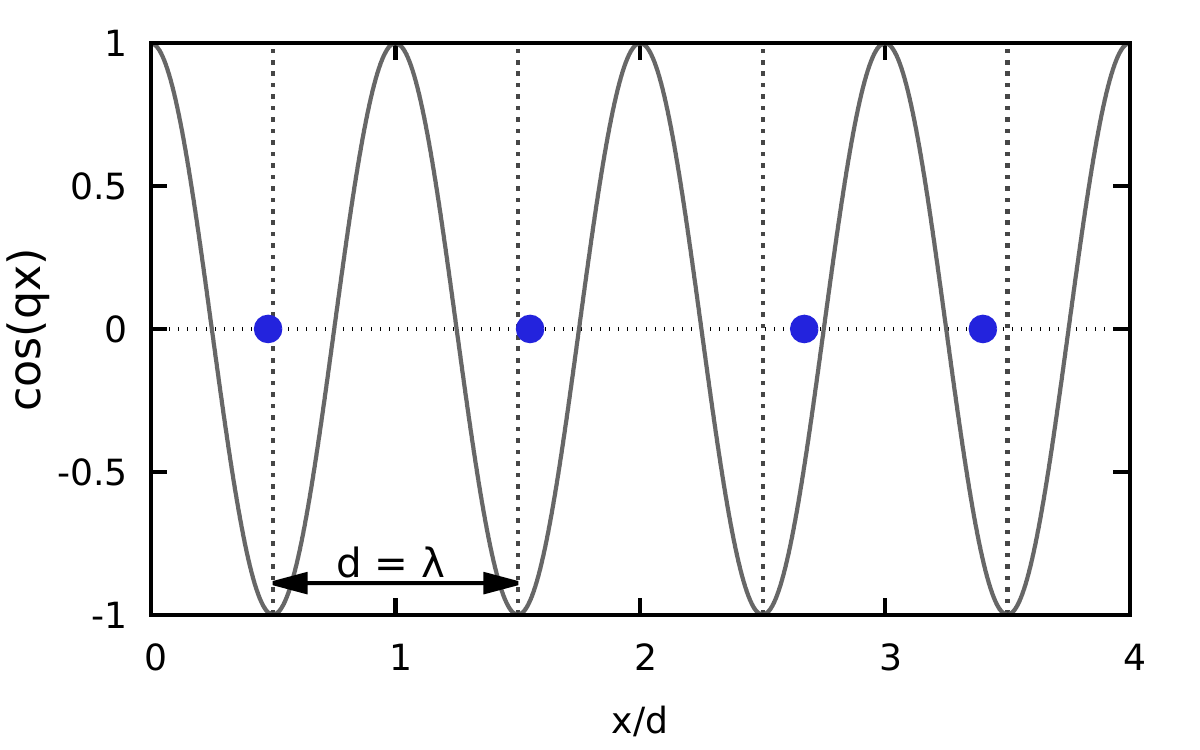}\\\vspace*{-0.94cm}
\includegraphics[width=0.49\textwidth]{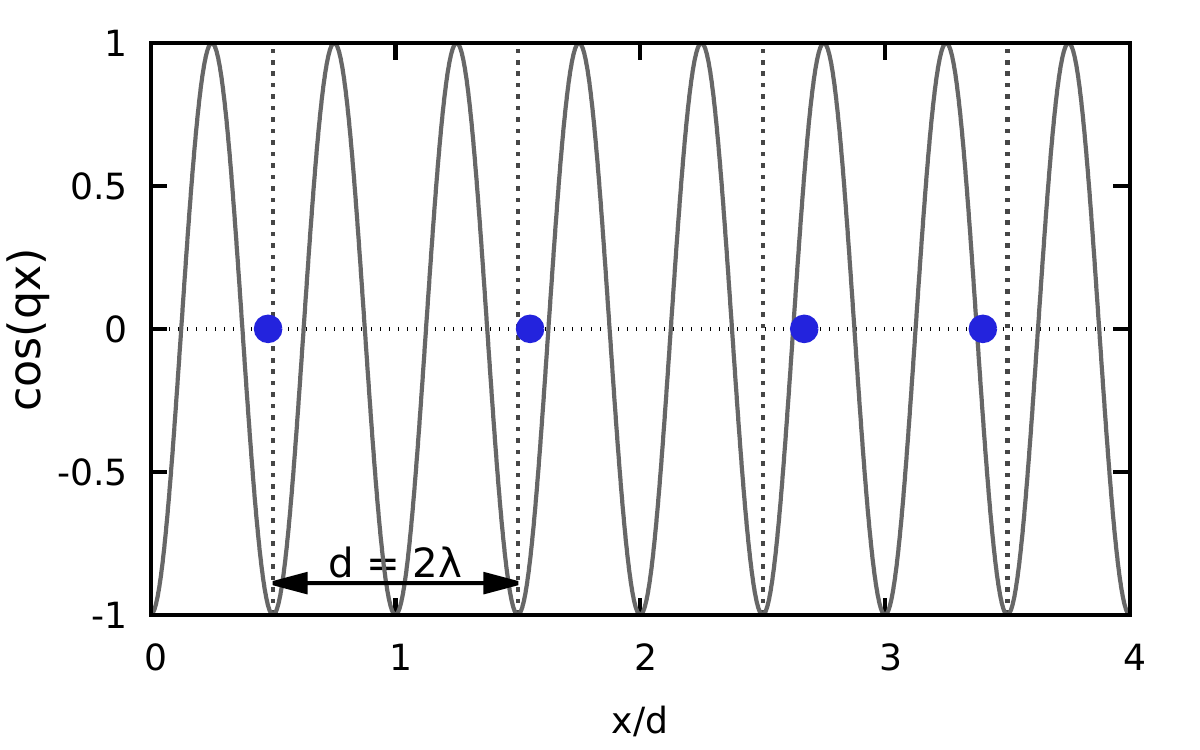}
\caption{\label{fig:explanation} Schematic illustration of the correlational origin of the first and second roton features. The wavelength $\lambda=2\pi/q$ of the first roton aligns with the average interparticle distance $d$, leading to a reduction in the average interaction energy of the system~\cite{Dornheim_Nature_2022} and a down-shift in the corresponding excitation energy. Reacting to a hypothetical external perturbation with half the wavelength (bottom) would trigger the same mechanism, although any deviation from the spatial arrangement with minimal energy leads to a more drastic reduction in the correlator compared to the first original roton.
}
\end{figure} 

As explained in detail in Ref.~\cite{Dornheim_Nature_2022}, understanding the physical origin of both roton features requires us to consider the involved length scales. In the top panel of Fig.~\ref{fig:explanation}, we illustrate a hypothetical configuration of four electrons (blue dots), which are approximately separated by the average interparticle distance $d$. It is easy to see that such a configuration would minimize the interaction energy $W$ and, thus, also maximize the static structure factor $S(\mathbf{q})$ when the wavelength $\lambda=2\pi/q$ is approximately equal to $d$.
At the same time, it is noted that the roton feature already clearly appears at $r_s\sim10$, when hardly any peak in $S(\mathbf{q})$ is visible. To resolve this apparent contradiction, we consider the inverse moment sum rule of the DSF~\cite{quantum_theory},
\begin{eqnarray}\label{eq:inverse_moment}
    \braket{\omega^{-1}} = \int_{-\infty}^\infty\textnormal{d}\omega\ \frac{S(\mathbf{q},\omega)}{\omega} = -\frac{2}{n} \chi(\mathbf{q},0)\ ,
\end{eqnarray}
which relates $S(\mathbf{q},\omega)$ to the static linear density response function $\chi(\mathbf{q},0)$. Eq.~(\ref{eq:inverse_moment}) immediately implies that the down-shift of spectral weight in $S(\mathbf{q},\omega)$ of the roton feature also manifests as a maximum in the density response, and vice versa. The reaction of the perturbed system will be maximal in absolute terms when the alignment to its minima coincides with a reduction in the interaction energy, as indeed the case for the roton; this is the basic idea behind the electronic pair alignment model proposed in Ref.~\cite{Dornheim_Nature_2022}, which explains both the existence of the roton feature despite the absence of a pronounced correlational peak in $S(\mathbf{q})$ as well as the exchange--correlation induced red-shift in $S(\mathbf{q},\omega)$ in the WDM regime.

\begin{figure}\centering
\includegraphics[width=0.49\textwidth]{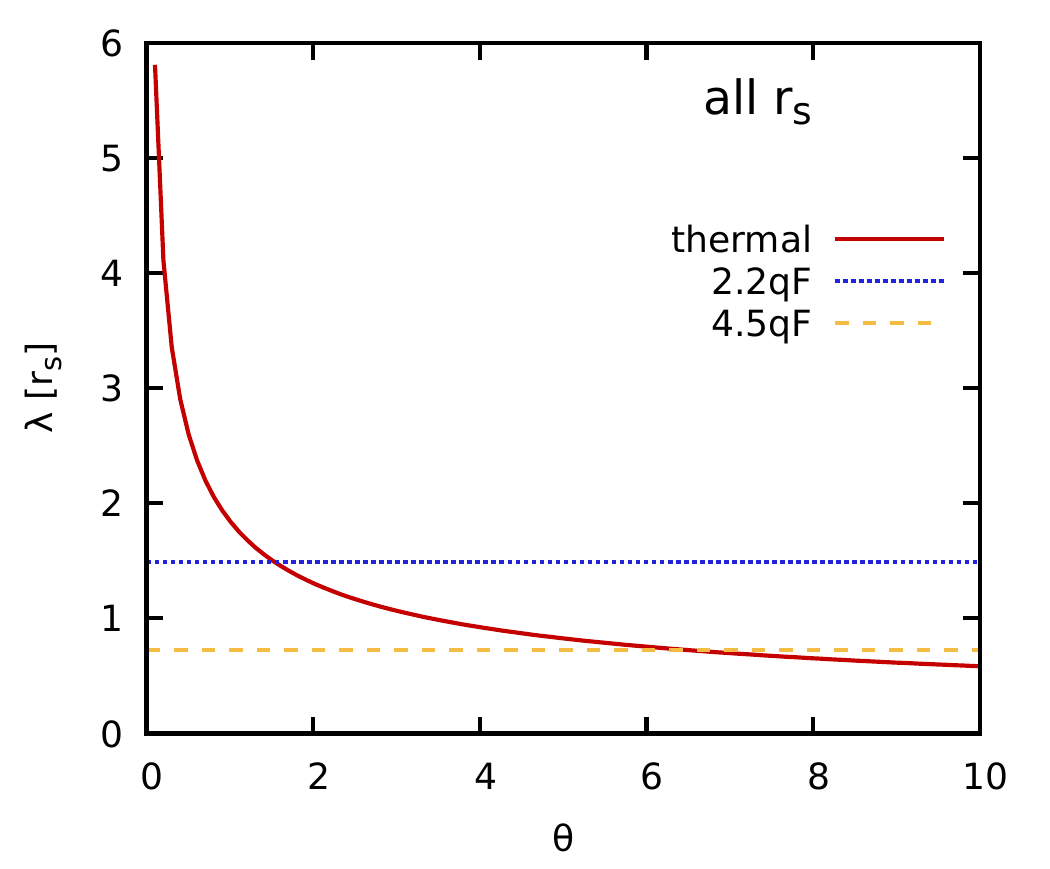}
\caption{\label{fig:ThermalWavelength} Comparison of the thermal de Broglie wavelength $\lambda=\sqrt{2\pi\beta}$ (solid red) to the wavelength of the first (dotted blue) and second roton (dashed yellow).
}
\end{figure} 

Obviously, the same reasoning applies for an external perturbation of twice the wavenumber and half the wavelength, i.e., the second harmonic of the original roton; see the bottom panel of Fig.~\ref{fig:explanation}.
At the same time, we would expect the magnitude of the second roton to be damped as the alignment with the cosinusoidal perturbation has to be even more perfect with the average interparticle distance to meaningfully contribute. 
A second damping effect, absent in the classical OCP is given by quantum delocalization. Specifically, we expect a reduction in the static linear density response and, hence, in the roton feature when the thermal wavelength $\lambda_\beta=\sqrt{2\pi\beta}$ is comparable to $\lambda=2\pi/q$. A corresponding comparison is shown in Fig.~\ref{fig:ThermalWavelength}, where $\lambda_\beta$ is shown as the solid red curve. Evidently, both roton features are expected to be damped by delocalization at $\Theta=1$, but the effect is substantially more severe at the larger wavenumber. Overall, we can interpret the second roton as an incipient phonon dispersion~\cite{Park_Nature_2024, Kalman_EPL_2010, Kalman_CPP_2012}, which will be more pronounced in an actual Wigner crystal.

We conclude the investigation of the coupling parameter dependence by inspecting the ITCF shown in Fig.~\ref{fig:ITCF}. The top row corresponds to the smallest possible wavenumber of $q=2\pi/L=0.64q_\textnormal{F}$, which has already been considered for selected densities in Fig.~\ref{fig:Large_Wavelength}. Evidently, the effect of the density is very pronounced at large wavelengths, and leads to a substantial reduction of the ITCF for higher $r_s$.
The central panel shows results for the ratio $F(\mathbf{q},\tau)/F(\mathbf{q},0)$, but the renormalization does not explain the observed trends; it does, however, conclusively rule out a constant factor in the DSF as their origin.
In the right panel, we have subtracted $F(\mathbf{q},0)$ instead of dividing by it. The small $\tau$-asymptotic is then governed by the same f-sum rule [Eq.~(\ref{eq:fsum})] asymptotic for all values of $r_s$, see the grey dotted line.
With increasing densities, the ITCFs converge towards the ideal Fermi gas result (dotted purple), whereas we observe a significant reduction in the $\tau$-decay in particular for $r_s\gtrsim100$. This nicely illustrates the interplay of Coulomb coupling and quantum delocalization in the strongly coupled regime.

The central row of Fig.~\ref{fig:ITCF} shows equivalent results for $q=2.08q_\textnormal{F}$, which is located in the vicinity of the original roton feature. Here, the ordering of the depicted ITCFs in the left panel is explained by the static structure factor $S(\mathbf{q})=F(\mathbf{q},0)$ depicted in Fig.~\ref{fig:SSF} above; it is more spiked for $r_s=300$, explaining the somewhat smaller value at this particular $q$ compared to $r_s=200$. Considering the ratio $F(\mathbf{q},\tau)/F(\mathbf{q},0)$ depicted in the central panel, we find that the UEG curves approach the ideal Fermi gas result with increasing density from above, which is in contrast to the smaller wavenumber depicted in the top row.
Again, this trend is already explained by the behavior of $S(\mathbf{q})$. Interestingly, the analysis of $F(\mathbf{q},\tau)-F(\mathbf{q},0)$, which is shown in the right panel, reveals that the absolute $\tau$-decay is smallest for $r_s=300$, which further attests to the pronounced roton feature at these conditions.

The bottom row of Fig.~\ref{fig:ITCF} shows the same information for a large wavenumber, $q=6.27q_\textnormal{F}$, which is close to the single-particle regime. Indeed, it holds $F(\mathbf{q},\beta/2)\approx0$ to a large degree for all simulated densities, leading to a convergence of $\Delta F_{\beta/2}(\mathbf{q})$ towards the ideal Fermi gas, see Fig.~\ref{fig:imaginary_dispersion} above. At the same time, we do observe small yet significant differences in the ITCFs around $\tau=\beta/10$ in all three panels. We thus conclude that, while often helpful to isolate certain effects, the consideration of a reduced property such as the relative $\tau$-decay does not necessarily replace the thorough investigation of the full ITCF and the rich physics encoded therein.

\begin{figure}\centering
\includegraphics[width=0.49\textwidth]{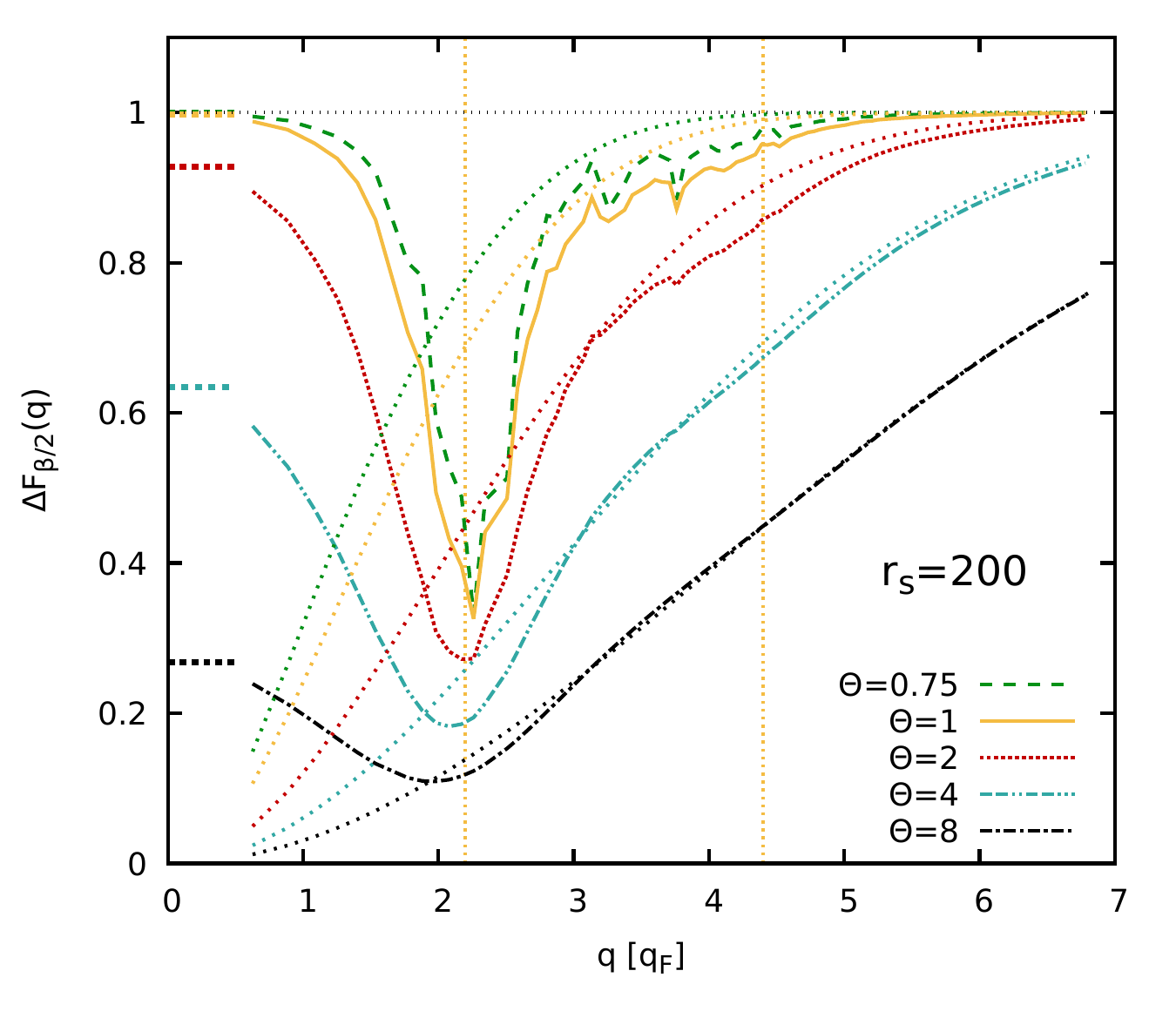}\\\vspace*{-1.05cm}\includegraphics[width=0.49\textwidth]{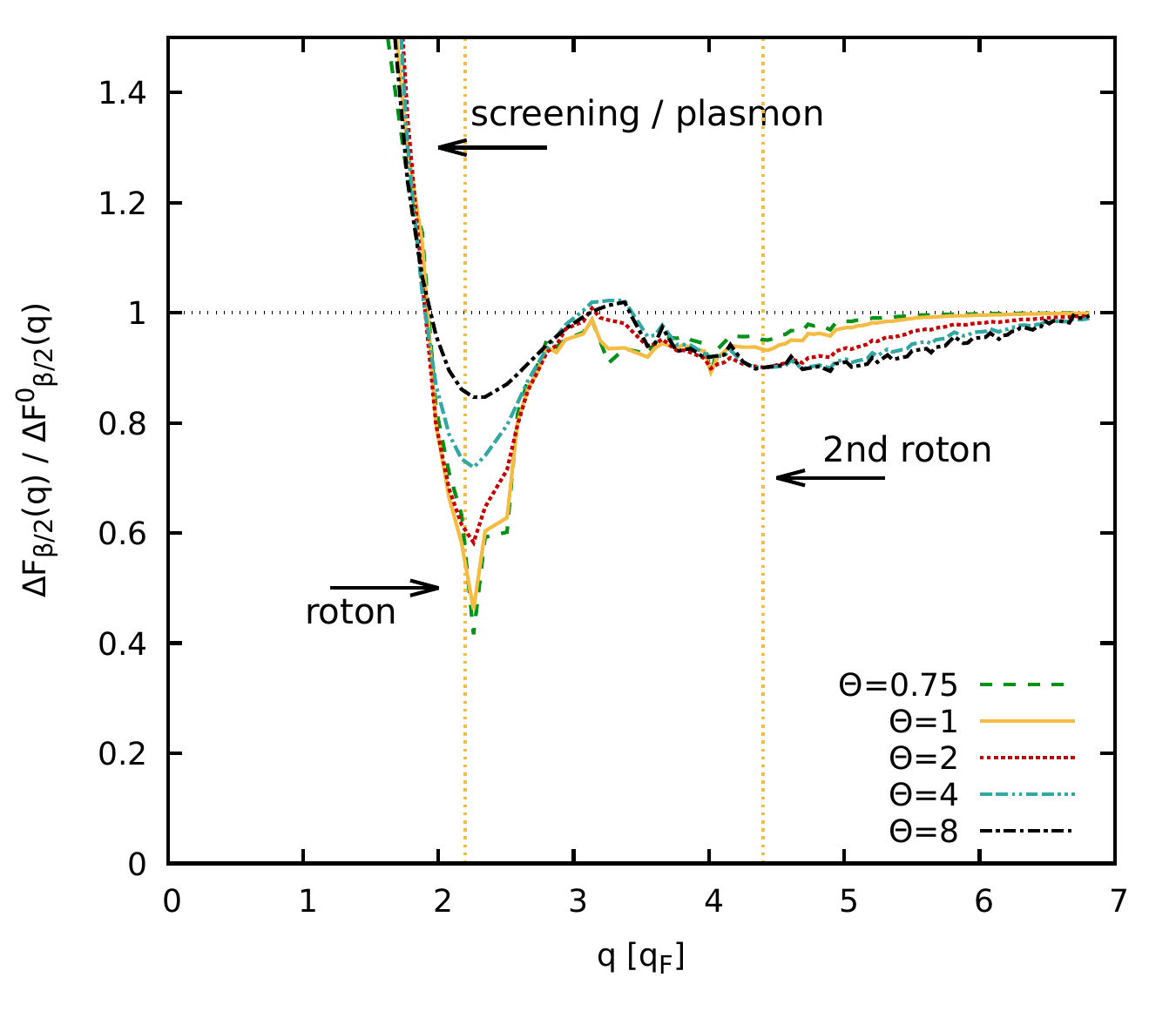}
\caption{\label{fig:rs200_theta} Top: \emph{ab initio} PIMC results for the temperature dependence of the relative $\tau$-decay $\Delta F_{\beta/2}(\mathbf{q})$ [Eq.~(\ref{eq:tau_decay})] of the ITCF at $r_s=200$. The horizontal bars at $q\to0$ correspond to the analytical plasmon limit, Eq.~(\ref{eq:imaginary_plasmon_beta_half}), the dotted lines have been included as a reference and show the $\tau$-decay of the ideal Fermi gas, the vertical yello lines indicate the wavenumber location of the first and second roton. Bottom: Same information, but normalized with respect to the ideal Fermi gas.
%\textcolor{red}{Relate thermal wavelength $\lambda_\Theta$ with wavelength of second roton, $\lambda=2\pi/q$.}
}
\end{figure} 

Let us conclude our imaginary time domain analysis by briefly considering the effect of the temperature. In the top panel of Fig.~\ref{fig:rs200_theta}, we show the relative $\tau$-decay, Eq.~(\ref{eq:tau_decay}), at $r_s=200$ for $0.75\leq\Theta\leq8$. Overall, we find pronounced temperature effects, that lead to smaller $\tau$-decay measures at higher temperatures. This is expected due to the smaller $\tau$-range that is available for the imaginary-time diffusion process at higher temperatures~\cite{Dornheim_MRE_2023,Dornheim_PTR_2023}. Furthermore, we find that the $\tau$-decay is described qualitatively by the ideal Fermi gas for $q\gtrsim3q_\textnormal{F}$ in particular for $\Theta=8$.
In the large wavelength limit, on the other hand, we find the usual finite plasmon limit (horizontal bars) for the UEG, whereas $\Delta F^0_{\beta/2}(\mathbf{q})$ (dotted curves) converges towards zero for all temperatures.

To properly resolve the first and second roton feature, we consider the ratio of the $\tau$-decay to the corresponding result for the ideal Fermi gas in the bottom panel of Fig.~\ref{fig:rs200_theta}. Thus, we can nicely resolve the first roton around the same wavenumber of $q\approx2.2q_\textnormal{F}$ for all the depicted temperatures; the depth of the first roton decreases with $\Theta$, as it is expected. In addition, we find a second roton at approximately twice the wavenumber (see also the vertical dotted yellow lines), with a depth that is considerably reduced compared to the original roton. This is has already been noted in the discussion of Fig.~\ref{fig:imaginary_dispersion} above; the depth of the second roton [when it is normalized with respect to the ideal Fermi gas, $\Delta F_{\beta/2}^0(\mathbf{q})$] becomes more pronounced with increasing temperatures. In fact, this seemingly counter intuitive observation can be explained by comparing the corresponding thermal wavelength $\lambda_\beta$ to the wavenumber of the roton, see Fig.~\ref{fig:ThermalWavelength}. Specifically, the second roton is strongly damped for lower temperatures, where $\lambda_\beta$ is substantially larger than $\lambda=2\pi/q$ for $q=4.5q_\textnormal{F}$. This situation changes with increasing $\Theta$, leading to a reduced impact of quantum delocalization and, in turn, a reduced damping of the second roton. Similar effects have been observed for the UEG interaction energy at intermediate temperatures~\cite{Groth_PRB_2016,Dornheim_PRB_2016}, and the plasmon frequency in isochorically heated aluminum~\cite{Moldabekov_PRR_2024}.

\subsection{Real-frequency perspective\label{sec:anal_cont}}

\begin{figure*}
    \centering
    \begin{minipage}[b]{0.685\textwidth}
        \includegraphics[width=\linewidth]{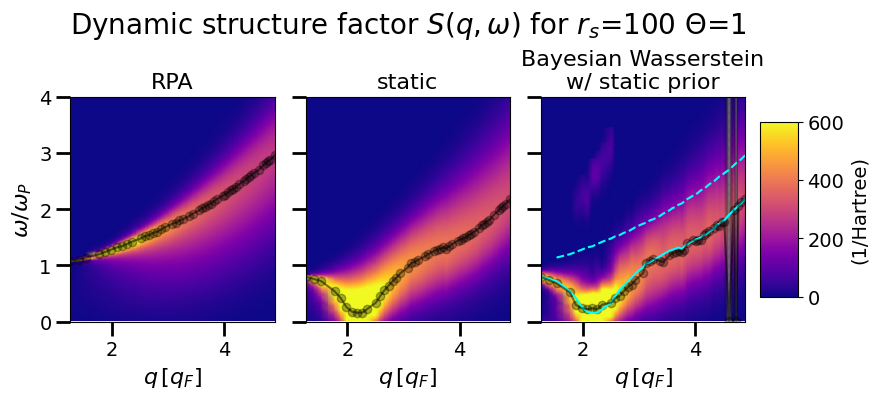}
        \includegraphics[width=\linewidth]{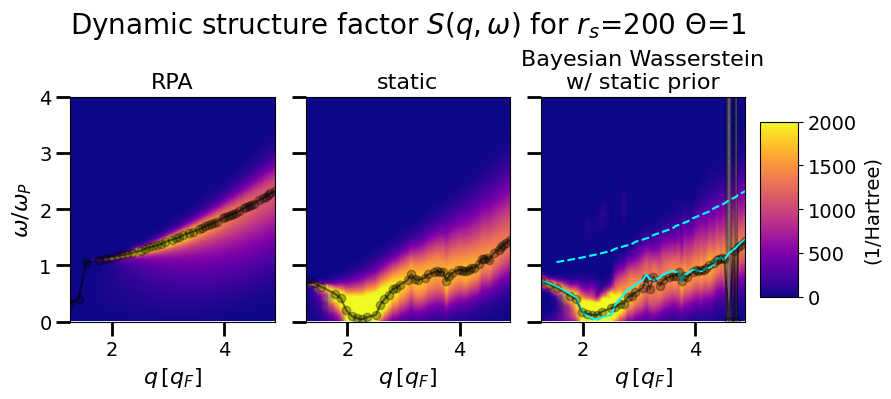}
    \end{minipage}%
    \begin{minipage}[b]{0.315\textwidth}
    \includegraphics[width=\linewidth]{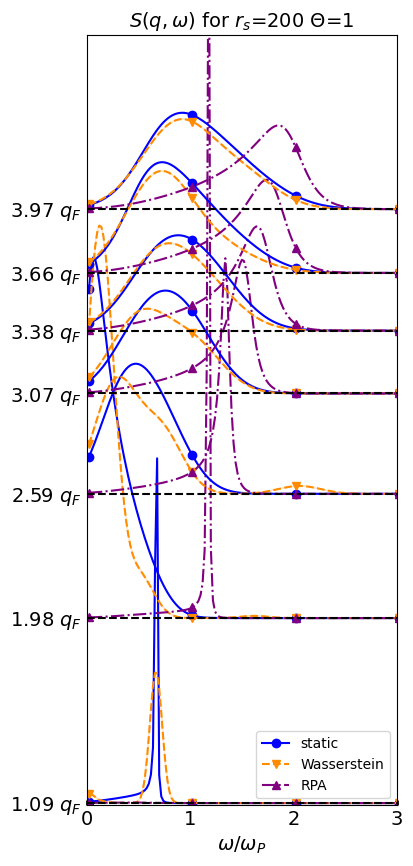}    
    \end{minipage}
    \caption{Left: Heat maps of the dynamic structure factor at $\Theta=1$ and $r_s=100$ (top), $r_s=200$ (bottom). The first DSF corresponds to the RPA, the second DSF corresponds to the static approximation, while the last DSF is estimated from ITCF data using PyLIT~\cite{pylit} informed by the static approximation Bayesian prior. The Bayesian label indicates that Gull's Bayesian posterior weighting~\cite{gull1989MEMBayesianWeighting} was used to average over possible solutions produced with the Wasserstein regularization. For both $r_s$ values, the analytic continuation yields a dispersion relation (marked by black dots) which is similar to the static approximation (marked by a solid cyan line) but exhibits large deviations from the RPA dispersion relation (marked by a dashed cyan line). Right: Fixed $q$ cross sections from the heat maps at $r_s=200$. PyLIT infers the dynamic dependence of the local field correction that the static approximation neglects, this alters the width and height of the peak.}
    \label{fig:AC-DSF}
\end{figure*}

\begin{figure}
    \centering
    \includegraphics[width=1\linewidth]{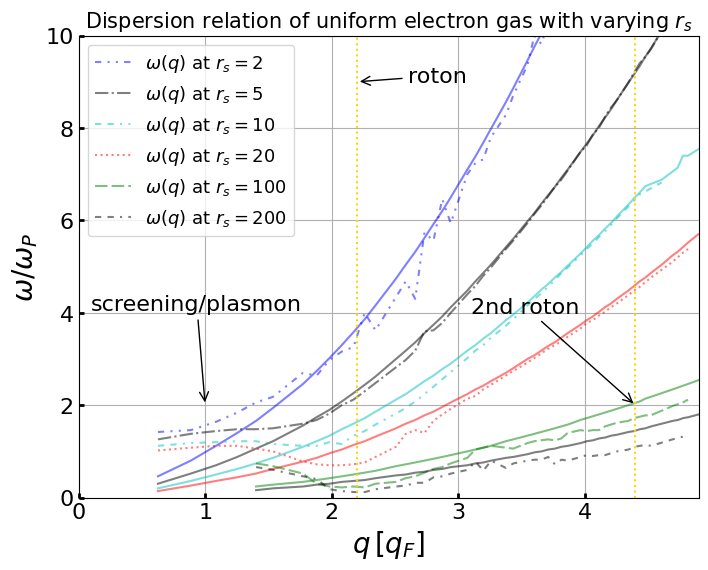}
    \includegraphics[width=1\linewidth]{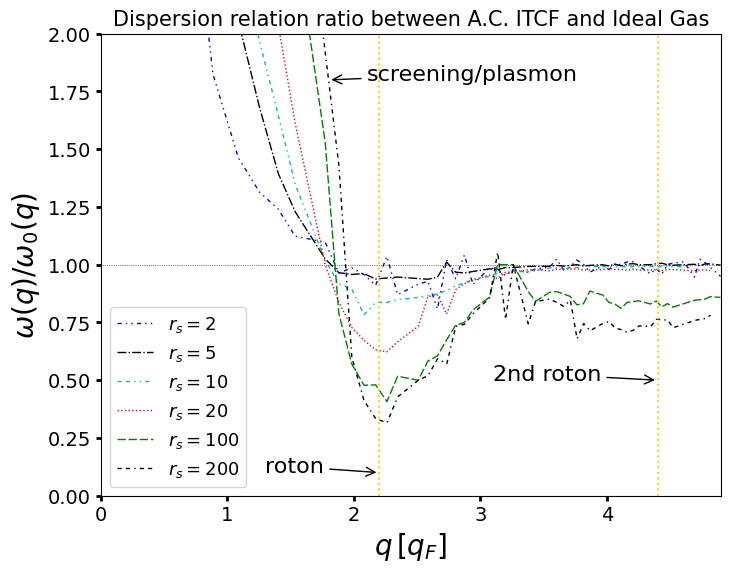}
    \caption{\label{fig:dispersion}Top: Plot of the dispersion relation varying $r_s$ for the Bayesian Wasserstein estimate of the DSF obtained via PyLIT ($\omega(q)$, dashed line). The single particle limit ($\omega_0(q)$) is given by solid lines. We see the emergence of a roton-like structure with $r_s \geq 10$ and the emergence of a second roton-like structure at $r_s=100$. Bottom: Plots of the dispersion relation estimated from analytic continuation $\omega(q)$ normalized by the ideal gas dispersion relation $\omega_0(q)$. This plot makes the distance between the dispersion relations evident to the reader. Further, these curves exhibit the same shape that was observed in Figure~\ref{fig:imaginary_dispersion}, with the second roton-like structure emerging at $r_s=100$ near $q =4.4 q_F$.}
    \label{fig:AC-DSF-DISPERSION}
\end{figure}

We continue our investigation of the dynamic properties of the electron liquid by switching to the more familiar frequency domain. For this purpose, we have carried out an analytic continuation to compute the dynamic structure factor $S(\mathbf{q},\omega)$ using the \texttt{PyLIT} code~\cite{pylit}; see Sec.~\ref{sec:anal_cont_theory} above and Ref.~\cite{pylit} for additional details. Note that the term dispersion relation is used in what follows as shorthand for the position of the maximum of the DSF, which coincides with the true dispersion (defined in terms of the roots of the dielectric function) only in the collective long wavelength limit~\cite{Hamann_CPP_2020}. It is important to point out again that the number of samples of $F(\mathbf{q},\tau)$ that are non-zero within the Monte Carlo error bars decreases with increasing $q$; see the bottom row of Fig.~\ref{fig:ITCF} for corresponding results at $q=6.27q_\textnormal{F}$. While PyLIT is uniquely positioned to handle this problem, we must still restrict ourselves to a $q$-range of $q\lesssim5q_\textnormal{F}$, where the analytic continuation is reliable.

The left panel of Fig.~\ref{fig:AC-DSF} features exemplary results for the DSF in the $q$-$\omega$-plane for $r_s=100$ (top) and $r_s=200$ (bottom) at $\Theta=1$. The first columns from the left correspond to the well known RPA, which converges towards the correct plasma frequency in the limit of $q\to0$, but which, otherwise, exhibits a simple $\omega(q)\sim q^2$ scaling of the position of the maximum of the DSF (black circles) corresponding to the single-particle dispersion relation. The second columns from the left have been computed within the \emph{static approximation} [see Sec.~\ref{sec:LRT} above] and substantially differ from the RPA. In contrast to the latter, the static approximation converges towards the plasma frequency from below, which is equivalent to the trends observed in the $\tau$-decay measure $\Delta F_{\beta/2}(\mathbf{q})$ shown in Figs.~\ref{fig:imaginary_dispersion},\ref{fig:theory_imaginary_dispersion}. Moreover, we find a pronounced roton feature at $q\approx2.2q_\textnormal{F}$ for both densities, followed by a visible feature (though not a minimum) at its second harmonic.  It is noted that the small wriggles in $\omega(q)$ at $r_s=200$ are a consequence of the aforementioned commensurability effects due to the finite size of the simulation box.

The third heatmap from the left correspond to our new analytic continuation results, where, as a reference, we have included the RPA and static approximation results for $\omega(q)$ as the dashed and solid cyan lines, respectively. In particular, the latter serves as the static approximation default model $D(\omega)$, which has been taken into account in the loss function, cf.~Sec.~\ref{sec:anal_cont_theory}, equations \eqref{eq:min_problem} and \eqref{eq:regularization}. Overall, we observe that the analytic continuation follows the default model to a large degree; a desirable outcome considering the already high quality of the static approximation in the description of $S(\mathbf{q},\omega)$~\cite{dornheim_dynamic,dynamic_folgepaper,chuna2025estimatesdynamicstructurefactor,pylit,Dornheim_Nature_2022}. 

Finally, the utmost right panel of Fig.~\ref{fig:AC-DSF} shows the $\omega$-dependence of the DSF at a selection of fixed $q$-values for the RPA, the static approximation and our AC results. For $q>2q_\textnormal{F}$, we find a more pronounced roton shift in the AC result compared to the static approximation and an incipient double peak structure, as it is expected~\cite{dornheim_dynamic, Dornheim_Nature_2022, chuna2025estimatesdynamicstructurefactor}. % For completeness, we note the small spectral feature at $q=2.59q_\textnormal{F}$ around twice the electronic plasma frequency. It is a spurious artifact caused by the employed Wasserstein regularization, see the topical discussion in Ref.~\cite{pylit}.

%At the same time, the Wasserstein and entropy priors predict a significantly deeper roton feature than the default model, which is in qualitative agreement with previous analytic continuation results at smaller values of $r_s$~\cite{dornheim_dynamic}; only the L2-norm basically reproduces the default model for the present choice of the regularization parameter.

In the top panel of Fig.~\ref{fig:AC-DSF-DISPERSION}, we show the dispersion of the DSF $\omega(q)$ for a host of densities at $\Theta=1$ computed using the analytic continuation. The solid lines show corresponding results for the single particle limit, $\omega_0(q)$, and have been included as a reference. We find the familiar trends with an incipient roton feature at $r_s\gtrsim10$, whereas $\omega(q)$ is basically indistinguishable from the single particle limit for $q\gtrsim3q_\textnormal{F}$ at $r_s=2$ and $r_s=5$. At the same time, the second roton is somewhat masked by the $\sim q^2$ increase in the dispersion even at $r_s=200$. Aiming to more clearly resolve the impact of exchange--correlation effects onto the dispersion relation of the UEG, we normalize $\omega(q)$ by $\omega_0(q)$ in the bottom panel of Fig.~\ref{fig:AC-DSF-DISPERSION}. In this representation, even the comparably small exchange--correlation induced red-shift at $r_s=2$ and $r_s=5$ at $q\approx2.2q_\textnormal{F}$ manifests as a shallow yet significant minimum. More important, we find a pronounced second roton at around twice the original wavenumber, which nicely substantiates the conclusions drawn from the relative $\tau$-decay measure reported in Sec.~\ref{sec:imaginary} above.

\section{Summary and Outlook\label{sec:outlook}}

We studied the dynamic properties of the finite temperature UEG over the broad range $2\leq r_s\leq 300$ based on highly accurate \emph{ab initio} PIMC simulations. Our analysis was carried out in the convenient imaginary--time as well as the conventional real frequency domain. \emph{First}, we directly employed the $\tau-$dependence of the ITCF $F(\mathbf{q},\tau)$ to study the density dependence of the roton-type feature that emerges at intermediate wavenumbers. As already hinted at in Ref.~\cite{Dornheim_MRE_2023}, the relative $\tau$-decay measure, Eq.~(\ref{eq:tau_decay}), is very well suited for such purpose, since it allows to resolve even subtle features in the dispersion relation. As expected, our analysis indicates that the roton becomes substantially more pronounced for stronger coupling. \emph{Second}, we employed the $\tau-$dependence of the ITCF to present clear evidence for the existence of a second roton at the second harmonic of the original roton feature for $r_s\gtrsim100$. \emph{Third}, from a physical perspective, we traced both roton features back to the minimization of the interaction energy for certain spatial patterns that align with roton wavenumbers. However, the second roton is damped both by deviations from perfect spatial ordering and by quantum delocalization effects. Therefore, its emergence can be interpreted as an incipient phonon type dispersion, which will become more pronounced in an actual Wigner crystal. \emph{Fourth}, we used our PIMC ITCF results at $r_s=200$ to study the effect of the temperature on both rotons, which turned out to be nontrivial. The original roton decreases in magnitude with increasing $\Theta$ (as expected), whereas the second roton becomes actually more pronounced for higher temperatures when normalized with respect to the ideal Fermi gas (which is rather counterintuitive). This has been explained by quantum delocalization induced damping, which is more severe at lower temperatures where the thermal wavelength $\lambda_\beta$ is larger. \emph{Finally}, we utilized our PIMC results for the ITCF as input for an analytic continuation to obtain the dynamic structure factor $S(\mathbf{q},\omega)$ using the recently published \texttt{PyLIT} code~\cite{pylit}. While being restricted to somewhat smaller wavenumbers, $q\lesssim5q_\textnormal{F}$, the real frequency domain results have fully confirmed the ITCF-based analysis of the first and second roton features, and also confirmed the trends observed in the previous Refs.~\cite{dornheim_dynamic,dynamic_folgepaper,chuna2025dualformulationmaximumentropy,chuna2025estimatesdynamicstructurefactor}.

In conclusion, we have established a rigorous dual-domain analysis of the complex interplay of quantum delocalization with Coulomb correlations in the finite temperature UEG spanning from the warm dense uniform electron gas to the strongly coupled electron liquid up to the vicinity of Wigner crystallization. All the PIMC results are freely available online and can serve as a thorough benchmark for other methodologies such as different schemes of the self-consistent dielectric formalism~\cite{tanaka_hnc,Tanaka_CPP_2017,castello2021classical,Tolias_JCP_2021,Tolias_JCP_2023,Tolias_PRB_2024,Tolias_CPP_2025}, different variants of the self-consistent method of moments~\cite{Arkhipov_2017,Arkhipov_2020,Filinov_PRB_2023,Filinov_RSTA_2023}, quantum versions of the quasi-localized charge approximation~\cite{Golden_2004QLCA,Golden_2006QLCA,Kahlert_2024} or molecular dynamics simulations with effective quantum pair potentials~\cite{Golubnichy_CPP_2002,Filinov_PRE_2004,Bonitz_POP_2024}. Future works might include the dedicated investigation of Wigner crystallization in the 3D UEG~\cite{Tozzini_1996,Drummond_PRB_Wigner_2004,Azadi_PRB_2022} and the corresponding analysis of the dispersion of the DSF maxima. This, however, will likely require simulations of $N\gtrsim10^3-10^4$ electrons, necessitating further optimizations such as an improved treatment of long-range interactions~\cite{Janke_PRX_2023} or the implementation of path contraction schemes~\cite{John_PRE_2016}. Future works might also include the analysis of the dispersion relation $\omega(q)$ and damping rate $\gamma(q)$ of collective excitations through the roots of the complex dielectric function within the weak damping rate approximation~\cite{bonitz_book} and beyond~\cite{Hamann_CPP_2020}.

In addition, we mention the enticing possibility to use PIMC simulations for the calculation of three-body and even higher-order imaginary-time correlation functions~\cite{Dornheim_JCP_ITCF_2021,Dornheim_CPP_2022,vorberger2024greensfunctionperspectivenonlinear}. First, higher-order ITCFs are directly related to the non-linear density response of the studied system, which is known to be even more sensitive to parameters such as the density and the temperature compared to the linear response~\cite{Dornheim_PRL_2020,Dornheim_PRR_2021}. Therefore, we expect a potentially increased sensitivity of higher-order ITCFs to incipient medium-range correlations and ordering, which might manifest similar to the roton. Second, many-body ITCFs are directly related to dynamic many-body properties such as the dynamic three-body structure factor~\cite{Dornheim_JCP_ITCF_2021,vorberger2024greensfunctionperspectivenonlinear}, which can, in principle, be computed by a future analytic continuation with respect to multiple imaginary-time dimensions. While undoubtedly difficult, the successful accomplishment of this task would open up the way to study dynamic many-body effects on a true \emph{ab initio} level. Finally, we note that, very recently, highly accurate PIMC results for the ITCF have become available also for light elements such as hydrogen~\cite{Dornheim_MRE_2024,Dornheim_JCP_2024,bellenbaum2025estimatingionizationstatescontinuum} and beryllium~\cite{Dornheim_JCP_2024,Dornheim_Science_2024,schwalbe2025density}. Subsequently, Moldabekov \emph{et al.}~\cite{moldabekov2025applyingliouvillelanczosmethodtimedependent} have demonstrated that linear-response time-dependent density functional theory is capable of accurately reproducing PIMC results for the ITCF in warm dense hydrogen.
Adapting the present analysis to the ITCF for those cases will thus open up new avenues to study dynamic effects in a gamut of materials, and might help to resolve the possible existence and experimental detectability of theoretically predicted phenomena such  as the roton feature in solid-density hydrogen~\cite{Hamann_PRR_2023}.

\appendix
\section{Exact long wavelength limit of ITCFs\label{sec:long_wavelength_derivation}}

\emph{Preliminaries.} The exact long-wavelength limit of the imaginary part of the density response function is given by\,\cite{pines_nozieresI_book,kugler_bounds}
\begin{equation}
\Im\{\chi(q\to0,\omega)\}=-\frac{\pi{n}q^2}{2m\omega_{\mathrm{pl}}}[\delta(\omega-\omega_{\mathrm{pl}})-\delta(\omega+\omega_{\mathrm{pl}})],\label{appendix:imDRFlong}
\end{equation}
with $\omega_{\mathrm{pl}}=\sqrt{4\pi{n}e^2/m}$ the plasma frequency. This expression demonstrates that the long wavelength UEG properties are entirely determined by undamped plasmons\,\cite{pines_nozieresI_book}, which is reflected on the fact that Eq.(\ref{appendix:imDRFlong}) alone exhausts the long-wavelength limit of the first and the third frequency moment sum rules for $\Im\{\chi(\boldsymbol{q},\omega)\}$\,\cite{kugler_bounds}. The expression also abides by the odd frequency parity of the imaginary part of the density response\,\cite{quantum_theory}. Combined with other exact expressions of linear response theory\,\cite{quantum_theory}, Eq.(\ref{appendix:imDRFlong}) serves as the generator of exact long-wavelength expressions\,\cite{kugler_bounds}. Substitution of Eq.(\ref{appendix:imDRFlong}) in the static limit of the Kramers-Kronig relations yields the long-wavelength limit of the static density response\,\cite{kugler_bounds}
\begin{equation}
\chi(q\to0)=-\frac{q^2}{4\pi{e}^2}.\label{appendix:SDRlong}
\end{equation}
Substitution of Eq.(\ref{appendix:imDRFlong}) in the Kramers-Kronig relations and its combination with the inverse dielectric function definition leads to the long-wavelength limit of the direct dielectric function\,\cite{pines_nozieresI_book}
\begin{equation}
\epsilon(q\to0,\omega)=1-\frac{\omega_{\mathrm{pl}}^2}{\omega^2}.\label{appendix:DDFlong}
\end{equation}
Substitution of Eq.(\ref{appendix:imDRFlong}) in the fluctuation--dissipation theorem leads to the long-wavelength limit of the dynamic structure factor (below written in a form that apparently abides by detailed balance),
\begin{align}
S(q\to0,\omega)&=+\frac{\hbar{q}^2}{2m\omega_{\mathrm{pl}}}\frac{1}{1-e^{-\beta\hbar\omega_{\mathrm{pl}}}}\left[\delta(\omega-\omega_{\mathrm{pl}})\right.\nonumber\\&\quad\quad\left.+e^{+\beta\hbar\omega}\delta(\omega+\omega_{\mathrm{pl}})\right]\,.\label{appendix:DSFlong}
\end{align}
Integration of Eq.(\ref{appendix:DSFlong}) over the entire real frequency domain yields the long-wavelength limit of the static structure factor\,\cite{review,kugler_bounds}
\begin{equation}
S(q\to0)=\frac{\hbar{q}^2}{2m\omega_{\mathrm{pl}}}\coth{\left(\frac{1}{2}\beta\hbar\omega_{\mathrm{pl}}\right)}.\label{appendix:SSFlong}
\end{equation}

\emph{A derivation in the imaginary--time domain}. The combination of the two-sided Laplace transform definition of ITCFs\,\cite{Dornheim_review,Tolias_JCP_2024} with the fluctuation--dissipation theorem yields
\begin{equation}
F(\boldsymbol{q},\tau)=-\frac{\hbar}{\pi{n}}\int_{-\infty}^{+\infty}\Im\{\chi(\boldsymbol{q},\omega)\}\frac{e^{-\hbar\omega\tau}}{1-e^{-\beta\hbar\omega}}d\omega\,.\label{appendix:FDT-ITCFcombo}
\end{equation}
Thus, the substitution of the exact long-wavelength limit of the imaginary part of the density response function, see Eq.(\ref{appendix:imDRFlong}), directly leads to the long-wavelength limit of the ITCF
\begin{equation}
F(q\to0,\tau)=\frac{\hbar{q}^2}{2m}\frac{1}{\omega_{\mathrm{pl}}}\left(\frac{e^{-\hbar\omega_{\mathrm{pl}}\tau}}{1-e^{-\beta\hbar\omega_{\mathrm{pl}}}}-\frac{e^{+\hbar\omega_{\mathrm{pl}}\tau}}{1-e^{+\beta\hbar\omega_{\mathrm{pl}}}}\right)\,.\nonumber
\end{equation}
After some straightforward exponential function algebra and the utilization of the hyperbolic sine addition formula, the long-wavelength limit of the ITCF can be ultimately rewritten in the more compact form
\begin{equation}
F(q\to0,\tau)=\frac{\hbar{q}^2}{2m\omega_{\mathrm{pl}}}\frac{\cosh{\left[\left(\tau-\frac{1}{2}\beta\right)\hbar\omega_{\mathrm{pl}}\right]}}{\sinh{\left(\frac{1}{2}\beta\hbar\omega_{\mathrm{pl}}\right)}}\,.\label{appendix:ITCFlong}
\end{equation}

\emph{A derivation in the Matsubara frequency domain}. The long-wavelength limit of the Matsubara density response $\widetilde{\chi}(\boldsymbol{q},\imath\omega_{\ell})$, where $\imath\omega_{\ell}=2\pi\imath\ell/\beta\hbar$ are the imaginary bosonic Matsubara frequencies, is first obtained. The Kramers-Kronig relation for imaginary frequency arguments stems from the consideration of the integral $\int_{\mathcal{C}}z\widetilde{\chi}(\boldsymbol{q},z)/(z^2+\omega_{\ell}^2)$ along the upper half plane contour consisting of the real axis and an infinite semi-circle. It reads as\,\cite{LandauLifshitz_SP1}
\begin{equation}
\widetilde{\chi}(\boldsymbol{q},\imath\omega_{\ell})=\frac{1}{\pi}\int_{-\infty}^{+\infty}\frac{\omega\Im\{\chi(\boldsymbol{q},\omega)\}}{\omega^2+\omega_{\ell}^2}d\omega\,.\label{appendix:KramersKronig}
\end{equation}
Thus, the substitution of the exact long-wavelength limit of the imaginary part of the density response function, see Eq.(\ref{appendix:imDRFlong}), directly leads to the long-wavelength limit of the Matsubara density response function
\begin{equation}
\widetilde{\chi}(q\to0,\imath\omega_{\ell})=-\frac{n}{m}q^2\frac{1}{\omega_{\mathrm{pl}}^2+\omega_{\ell}^2}\,.\label{appendix:MatsubaraDRFlong}
\end{equation}
Note that for $\ell=0$, Eq.(\ref{appendix:MatsubaraDRFlong}) collapses to Eq.(\ref{appendix:SDRlong}). The exact Fourier-Matsubara series representation of the ITCF reads as\,\cite{Tolias_JCP_2024}
\begin{equation}
F(\boldsymbol{q},\tau)=-\frac{1}{n\beta}\sum_{\ell=-\infty}^{+\infty}\widetilde{\chi}(\boldsymbol{q},\imath\omega_{\ell})e^{-\imath\hbar\omega_{\ell}\tau}\,.\label{appendix:MatsubaraFourierSeries}
\end{equation}
The combination of Eqs.(\ref{appendix:MatsubaraDRFlong},\ref{appendix:MatsubaraFourierSeries}) yields
\begin{equation}
F(q\to0,\tau)=\frac{1}{m\beta}q^2\left(\frac{\beta\hbar}{2\pi}\right)^2\sum_{\ell=-\infty}^{+\infty}\frac{\exp{\left(-\imath\ell\frac{2\pi}{\beta}\tau\right)}}{\ell^2+\left(\frac{\beta\hbar}{2\pi}\omega_{\mathrm{pl}}\right)^2}\,.\nonumber
\end{equation}
The derivation now requires use of the tabulated infinite series $\sum_{n=-\infty}^{+\infty}e^{-\imath{n\alpha}}/[n^2+\beta^2]=(\pi/\beta)\{\sinh{(\beta\alpha)}+\sinh{[\beta(2\pi-\alpha)]}\}/[\cosh{(2\pi\beta)}-1]$, which holds for $0\leq\alpha\leq2\pi$\,\cite{GradshteynRyzhik}. The latter inequality is respected courtesy of $0\leq\tau\leq\beta$. Substituting, one acquires
\begin{equation}
F(q\to0,\tau)=\frac{\hbar{q}^2}{2m\omega_{\mathrm{pl}}}\frac{\sinh{\left(\hbar\omega_{\mathrm{pl}}\tau\right)}+\sinh{\left[\left(\beta-\tau\right)\hbar\omega_{\mathrm{pl}}\right]}}{\cosh{\left(\beta\hbar\omega_{\mathrm{pl}}\right)}-1}\,.\nonumber
\end{equation}
Application of the hyperbolic sine addition formula and the hyperbolic cosine double angle formula, ultimately leads to the same compact form
\begin{equation}
F(q\to0,\tau)=\frac{\hbar{q}^2}{2m\omega_{\mathrm{pl}}}\frac{\cosh{\left[\left(\tau-\frac{1}{2}\beta\right)\hbar\omega_{\mathrm{pl}}\right]}}{\sinh{\left(\frac{1}{2}\beta\hbar\omega_{\mathrm{pl}}\right)}}\,.\label{appendix:ITCFlongagain}
\end{equation}

\emph{Further cross-checks}. The compact form of the long-wavelength limit of the ITCF automatically satisfies the imaginary time translation invariance $F(\boldsymbol{q},\tau)=F(\boldsymbol{q},\beta-\tau)$ and the normalization $F(\boldsymbol{q},0)=S(\boldsymbol{q})$ properties of ITCFs\,\cite{Dornheim_T_2022,Tolias_CtPP2025}. As a validation test, it is easy to verify that the long-wavelength limits of the static density response and the ITCF, see Eqs.(\ref{appendix:SDRlong},\ref{appendix:ITCFlong}), are indeed connected via the imaginary time version of the fluctuation--dissipation theorem\,\cite{Sugiyama_PRB_1994,Dornheim_MRE_2023,Dornheim_PTR_2023}, which holds for any wavenumber and has been discussed in the main text,
\begin{equation}
\chi(\boldsymbol{q})=-n\int_0^{\beta}F(\boldsymbol{q},\tau)d\tau.
\end{equation}

\begin{acknowledgements}

\noindent This work was partially supported by the Center for Advanced Systems Understanding (CASUS), financed by Germany’s Federal Ministry of Education and Research and the Saxon state government out of the State budget approved by the Saxon State Parliament. This work has received funding from the European Union's Just Transition Fund (JTF) within the project \emph{R\"ontgenlaser-Optimierung der Laserfusion} (ROLF), contract number 5086999001, co-financed by the Saxon state government out of the State budget approved by the Saxon State Parliament. This work has received funding from the European Research Council (ERC) under the European Union’s Horizon 2022 research and innovation programme (Grant agreement No. 101076233, "PREXTREME"). Views and opinions expressed are however those of the authors only and do not necessarily reflect those of the European Union or the European Research Council Executive Agency. Neither the European Union nor the granting authority can be held responsible for them. Computations were performed on a Bull Cluster at the Center for Information Services and High-Performance Computing (ZIH) at Technische Universit\"at Dresden and at the Norddeutscher Verbund f\"ur Hoch- und H\"ochstleistungsrechnen (HLRN) under grant mvp00024.
\end{acknowledgements}

%%%%%%%%%%%%%%%%%%%%%%%%%%%%%%%%%%%%%%%%%%%%%%%%%%%%%%%%%%%%%%%%%%%%%%%%%%%%%%%%
% literature
%%%%%%%%%%%%%%%%%%%%%%%%%%%%%%%%%%%%%%%%%%%%%%%%%%%%%%%%%%%%%%%%%%%%%%%%%%%%%%%%
\bibliography{bibliography}

\begin{thebibliography}{190}
\bibitem [{rep()}]{repo}%
  \BibitemOpen
  \bibinfo {note} {A link to a repository containing all PIMC results will be made available upon publication.}\BibitemShut {Stop}%
\end{thebibliography}
\end{document}